\documentclass[%
 reprint,
 amsmath,amssymb,
 aps,
 onecolumn,
]{revtex4-1}

\usepackage{xcolor}
\usepackage{graphicx}
\usepackage{dcolumn}
\usepackage{bm}
\usepackage[autopunct=true]{csquotes}
\DeclareMathOperator*{\argmin}{arg\,min}
\def\lim{{\rm lim}}

\DeclareMathOperator{\extr}{extr}
\DeclareMathOperator{\erfc}{erfc}

\newcommand{\lp}{\ell_p}
\newcommand{\lone}{\ell_1}
\newcommand{\leps}{\ell_\epsilon}
\newcommand{\lzero}{\ell_0}
\newcommand{\bxo}{\bm{x}_o}
\newcommand{\rhoo}{\rho_o}
\newcommand{\bF}{\bm{F}}
\newcommand{\by}{\bm{y}}
\newcommand{\bx}{\bm{x}}
\newcommand{\bhx}{\bm{\hat{x}}}
\newcommand{\R}{{\rm I\!R}}
\newcommand{\E}{{\rm I\!E}}
\newcommand{\N}{{\rm I\!N}}
\newcommand{\ansatz}{\emph{Ansatz} }

\newcommand{\diveta}{\mathrm{div}\,\eta}

\newcommand{\mASPo}{{\mathrm{ASP}_o}}

\def\BibTeX{{\rm B\kern-.05em{\sc i\kern-.025em b}\kern-.08em
    T\kern-.1667em\lower.7ex\hbox{E}\kern-.125emX}}

\begin{document}

\preprint{APS/123-QED}

\title{The phase diagram of compressed sensing with  $\ell_0$-norm regularization}%

\author{Damien Barbier}
\affiliation{%
\'Ecole Polytechnique Fédérale de Lausanne, Information, Learning and Physics lab and Statistical Physics of Computation lab, CH-1015 Lausanne, Switzerland}

\author{Carlo Lucibello}

\author{Luca Saglietti}
\affiliation{
Institute for Data Science and Analytics, Bocconi University, Milan, Italy}%
\affiliation{
Department of Computing Sciences, Bocconi University, Milan, Italy}%

\author{Florent Krzakala}
\affiliation{%
\'Ecole Polytechnique Fédérale de Lausanne, Information, Learning and Physics lab and Statistical Physics of Computation lab, CH-1015 Lausanne, Switzerland }

\author{Lenka Zdeborov\'a }
\affiliation{%
\'Ecole Polytechnique Fédérale de Lausanne, Statistical Physics of Computation Laboratory, CH-1015 Lausanne, Switzerland}%

\begin{abstract}
Noiseless compressive sensing is a two-steps setting that allows for undersampling a sparse signal and then reconstructing it without loss of information. The LASSO algorithm, based on $\lone$ regularization, provides an efficient and robust to address this problem, but it fails in the regime of very high compression rate. Here we present two algorithms based on $\lzero$-norm regularization instead that outperform the LASSO in terms of compression rate in the Gaussian design setting for measurement matrix. These algorithms are based on the Approximate Survey Propagation, an algorithmic family within the Approximate Message Passing  class. In the large system limit, they can be rigorously tracked  through State Evolution equations and it is possible to exactly predict the range compression rates  for which perfect  signal reconstruction is possible. We also provide a statistical physics analysis of the $\lzero$-norm noiseless compressive sensing model. We show the existence of both a replica symmetric state and a 1-step replica symmmetry broken (1RSB) state for sufficiently low $\lzero$-norm regularization. The recovery limits of our algorithms are linked to the behavior of the 1RSB solution.
\end{abstract}

\maketitle

\section{Introduction}
\label{sec: introduction}
 Compressed (or compressive) sensing (CS) is an approach that enables perfect recovery of signals (e.g. images) with far fewer measurements than the signal dimensionality. In order to achieve this, 
 CS relies on the fact that encoded signals usually contain redundancies, which implies that given a well-chosen basis the signal becomes sparse (i.e. contains many zero entries). This situation appears for example in magnetic resonance imaging, where the process can be sped up by minimizing the number of Fourier measurements needed to reconstruct an image frame. 
 More practically, CS involves the reconstruction of a sparse $N$-dimensional vector $\bxo$, the signal, from a $M$-dimensional feature vector $\by$, whose components represent a set of measurements on the signal. The measurements are obtained via a linear transformation acting on the signal using a $M\times N$ matrix $\bF$, i.e. $\by=\bF \bxo$. In this setup, the feature vector and the measurement matrix constitute the only accessible information. To account for the compression of information, the number of measurements $M$ is taken smaller than the signal dimension $N$. When $M<N$ the system of linear equations is under-determined and the signal cannot be recovered by inverting the matrix $\bF$. Theoretically, if the signal contains $\rhoo N$ (with $0\leq \rhoo\leq 1$) non-zero entries, $\bxo$ could be recovered exactly as long as $M>\rhoo N$ but this would require the observer to know which entries of the signal are equal to zero. Therefore the aim of any realistic recovery algorithm is to obtain the best compression rate $\alpha_c(\rhoo)=M/N$ given that $\rhoo\leq \alpha_c(\rhoo)\leq 1$.

 Many theoretical works have been inspired by this problem \cite{Donoho2006,Candes2006,Kabashima2009,Ganguli2010,Krzakala2012_1,Krzakala2012_2} and several algorithms have been proposed to recover the true signal. Of particular interest for us are the statistical physics-inspired works including message-passing algorithms \cite{L2009,Bayati2011}, Bayes-optimal statistics \cite{Krzakala2012_1,Krzakala2012_2} and $\lp$-norm denoisers \cite{Kabashima2009,Donoho2011,Xu2013,Zheng2017} 
 among others \cite{Figueiredo07,Kabashima2010,Obuchi2018,Bora2017}.
 In this paper, we propose to use a $\lzero$-norm penalty to enforce the sparsity in the reconstructed signal. Previous works have already tried this approach using statistical physics methods \cite{Bereyhi2017,Zheng2017,Kabashima2010}, however many difficulties, mostly induced by the discontinuity of the $\lzero$-norm, have prevented a clear understanding of this setting. 
 In the following, we fully characterize the properties of the minimizers of an $\lzero$-regularized cost function within a statistical physics framework. 
 We also provide two novel algorithmic schemes, closely related to and well-described by our theoretical analysis. The algorithms belong to the approximate message passing family \cite{L2009}, and more specifically are instances of survey propagation \cite{Antenucci2019, Lucibello19a}.
 
The main results presented in this paper are of two distinct natures. 
First, we derive two message-passing algorithms that can be rigorously tracked via State Evolution (SE) \cite{Bayati2011} and that outperform LASSO in the signal reconstruction task. 
Then, we present a replica analysis of the CS inference problem, which is shown to be closely related to SE for the presented algorithms. 
We highlight the coexistence in the free energy of two stable minima, respectively described by an RS and a 1RSB ansatz: the first is an informed global minimum in strongly aligned to the signal, while the second is a fixed point that describes the $\lzero$-norm signal-recovery protocol that can be reached from an uninformed initialization. Moreover, we show that the 1RSB fixed point is stable towards further replica symmetry breaking for sufficiently low regularization. 

\section{The model}
\label{sec: the model}
We assume an $N$-dimensional signal vector $\bxo \in \R^N$, compressed to a $M$-dimensional vector $\by \in \R^M$ (with $M\leq N$) via the protocol
$ \by=\bF\bxo$ ,
where $\bF\in \R^{M\times N}$ is the so-called measurement matrix. We assume that all components $F_{ij}$ are i.i.d variables  $F_{ij}\sim \mathcal{N}(0, N^{-1})$. Additionally, the entries ${x_{o,i}}$ are independently drawn from a Gauss-Bernoulli distribution
\begin{align}
\label{eq: signal distrib}
P_o(x_{o,i})
\!=\!(1-\rhoo)\delta(x_{o,i})+\rhoo \frac{e^{-\frac{1}{2}x^2_{o,i}}}{\sqrt{2\pi}},
\end{align}
with $0\leq \rhoo\leq 1 $. Consequently, the average number of non-zero entries in $\bxo$ is $\rhoo N$.
To recover the signal $\bxo$ given  the feature vector $\by$ and the measurement matrix $\bF$, our approach will be to solve the minimization problem
\begin{align}
\label{eq: minimization formulation}
\bx^\star=\argmin_{\bx} \Vert\by-\bF\bx\Vert_2^2+\lambda \Vert \bx \Vert_0, 
\end{align}
for some regularization parameter $\lambda  >0$, eventually to be sent to zero. 
We recall that the $\lzero$-norm is defined as
\begin{align}
\Vert \bx \Vert_0=\sum_i |x_{i} |^0 ;\qquad \;\; |x_{i}|^0 =\begin{cases}
       1 & \text{if}\;\, x_{i}\neq 0, \\
       0 &  \text{if}\;\, x_{i}= 0. 
    \end{cases}
\end{align}
The first term in the r.h.s of Eq.~\eqref{eq: minimization formulation} accounts for the measurement protocol: if a configuration $\bx$ verifies $\by=\bm{Fx}$ this term will be null. 
Then, the role of the $\lzero$-norm penalty in Eq.~\eqref{eq: minimization formulation} is to bias the system towards configurations with a certain level of sparsity, controlled by the magnitude of the coefficient $\lambda$. Previous works have already addressed the possibility of obtaining new algorithms with a $\lzero$-norm penalty that could outperform pre-existing procedures (like LASSO) \cite{Bereyhi2019,Kabashima2009,Kabashima2010}.

In the rest of this paper, we will focus on the asymptotic case $N\rightarrow +\infty$ with $M/N=\alpha$ remaining finite. Under these assumptions, the optimization problem can be reformulated as a statistical physics problem where $\bx^\star$ is obtained from sampling the probability distribution
\begin{align}
\label{eq: sampling probability distrib}
P_{\lambda,\epsilon,\beta}(\bx)=\frac{1}{Z_{\lambda,\epsilon,\beta}}e^{-\beta\mathcal{L}_{\lambda,\epsilon}(\bx)},
\end{align}
where $\beta$ is an inverse temperature to be sent to infinity (zero temperature limit), and the partition function $Z_{\lambda,\epsilon,\beta}$ and the cost functions $\mathcal{L}_{\lambda,\epsilon}$ are respectively given by
\begin{align}
\label{eq: partition function}
Z_{\lambda,\epsilon,\beta} &=\int d\bx\;e^{-\beta\mathcal{L}_{\lambda,\epsilon}(\bx)}\, ,\\
\mathcal{L}_{\lambda,\epsilon}(\bx)&=\Vert\by-\bF\bx\Vert_2^2+\lambda\lVert \bx \rVert^\epsilon_\epsilon\; .
\end{align}
The extremizers of Eq.~\eqref{eq: minimization formulation} are recovered in the limit where first the inverse temperature $\beta$ is sent to $+\infty$ and then $\epsilon\to {0^+}$.
The necessity to consider in the intermediate calculations before the zero temperature limit an $\leps$-norm, $\lVert \bx \rVert^\epsilon_\epsilon= \sum_i |x_i|^\epsilon$ , stems from the fact that in Eq.~\eqref{eq: partition function} the $\lzero$-norm is constant almost everywhere. 

For convenience, we define $P_\lambda(\bx)= \lim_{\epsilon\to {0^+}}\,\lim_{\beta\to\infty}\,P_{\epsilon,\beta}(\bx)$, where the ordering of the limits is relevant. Similarly, we define $\mathcal{L}_\lambda(\bx) =\lim_{\epsilon\to 0^+} \mathcal{L}_{\lambda,\epsilon}(\bx)$.
We denote with $\langle \cdot \rangle_\lambda$ expectations with respect to $P_\lambda(\bx)$, and decompose the expected loss function in a data-dependent term and a density term as follows:
\begin{align}
&\frac{1}{N}\left\langle\mathcal{L}_\lambda(\bx) \right\rangle_\lambda =e_\lambda +\lambda\,\rho_\lambda\\
\text{with}\quad &e_\lambda=\frac{1}{N}\left\langle \Vert\by-\bF\bx\Vert_2^2 \right\rangle_\lambda\quad{\rm and}\quad \rho_\lambda=\frac{1}{N}\left\langle\Vert \bx \Vert_0\right\rangle_\lambda.
\end{align}
While once taken the limits $\beta\to+\infty$ followed by $\epsilon\to 0^+$, the distribution $P_\lambda(\bx)$ we obtain concentrates on the solutions of problem \eqref{eq: minimization formulation}, in order to recover the true signal $\bxo$ we also have to consider an annealing procedure for $\lambda$. In the successive limit $\lambda\to 0^+$ we will obtain an estimation of the true signal as the minimum-norm interpolator with zero data loss (i.e. $e=0$).

 \section{Algorithm for $l_0$-norm compressed sensing}
\label{sec: algorithms for CS}
In Ref. \cite{Candes2006} it has been proven that, as long as long as the number of observations is larger than the number of non-zero entries in the true-signal, that is $\alpha>\rho_o$, the probability distribution $P_\lambda$ concentrates onto the signal, i.e.
\begin{align}
  \lim_{\lambda\to 0^+}P_\lambda(\bx)=\delta(\bx-\bxo)\, .
\end{align}

In our approach, we aim to efficiently estimate the marginal $\hat{\bx}= \langle \bx \rangle_\lambda$ for a sequence of $\lambda$ values that go to zero in order to solve the problem.
In this Section, we will address the two related questions for the large $N$ limit of our synthetic setting: can we estimate efficiently (polynomial time) the marginals of $P_{\lambda}(x)$? Do the algorithms we derive recover the signal $\bxo$ and in which region of the parameters $\alpha$ and $\rho_o$? The answer to these questions will be linked to the structure of the free energy landscape discussion in Sec.~\ref{sec:stat mech analysis}.

\subsection{Hard-thresholding algorithms}
\label{subsec: HT algorithms}
Here we focus on two algorithms, one based on Approximate Message Passing (AMP)\cite{L2009}  and one based on proximal gradient descent. We will find that the AMP algorithm, normally highly effective in solving generalized linear models, completely fails in the $\lzero$-reconstuction problem. The same goes for the proximal gradient algorithms. Nevertheless, they will provide a stepping stone for the effective message passing algorithms presented in Section \ref{subsec: ASP and ASPo}.   Both algorithms rely on the hard-thresholding denoising function

\begin{eqnarray}
\eta_{\rm HT}(u,\lambda)=\left\{
    \begin{array}{ll}
       0 & \text{if}\; \vert u \vert<\sqrt{2\lambda} \\
       u & \text{if}\; \vert u \vert>\sqrt{2\lambda} \\
    \end{array}
\right. \; .
\end{eqnarray}
Below, we will use the shorthand notation ${\eta}_{\rm HT}' (u,\lambda)=\partial_u \eta_{\rm HT}(u,\lambda)$.

Under our statistical setting, AMP allows for the estimation of the expectation $\langle \bx \rangle_{\lambda}$ over the graphical model $P_\lambda$. The derivation of the AMP equations is quite standard \cite{Krzakala2012_1,Montanari2010}, and has been extensively discussed for the family of generalized linear models to which our problem belongs. Once derived the message passing equation for finite $\beta$ and $\epsilon >0$, some caution and appropriate rescalings are needed in order to take the limits $\beta\rightarrow+\infty$ followed by $\epsilon\to0^+$ in the message passing rules (see App.~\ref{app: finite temp algo} for a discussion). After the limits, setting the initial condition to some $\bm{\hat{x}}^{t=0}$ (e.g. zeros or standard Gaussian) along with $\bm{{z}}^{t=0}=\bm{y}$, $A^{t=0}=\alpha$, the AMP iterations for $t\geq 1$ read  
\begin{align}
\label{eq: AMP algo}
\bm{\hat{x}}^{t}&=\eta_{\rm HT}\left(\frac{\bm{F}^{\intercal}\bm{z}^{t-1}}{\alpha}+\bm{\hat{x}}^{t-1},\frac{\lambda}{{A}^{t-1}}\right)\, ,\\
\bm{z}^t&=\bm{y}\!-\!\bm{F}\bm{\hat{x}}^t \!+\!\frac{\bm{z}^{t-1}}{\alpha} \diveta_{\rm HT} \left(\frac{\bm{F}^{\intercal}\bm{z}^{t-1}}{\alpha}\!+\bm{\hat{x}}^{t-1},\frac{\lambda}{{A}^{t-1}}\right)\!\!\, ,\nonumber\\
A^{t}&={\alpha}\Bigg[1+\frac{\diveta_{\rm HT} \left(\frac{\bm{F}^{\intercal}\bm{z}^{t-1}}{\alpha}+\bm{\hat{x}}^{t-1},\frac{\lambda}{{A}^{t-1}}\right)}{A^{t-1}}\Bigg]^{-1}\, ,\nonumber
\end{align}
with
\begin{eqnarray}
 \diveta_{\rm HT}(\bm{u},\lambda)&=&\frac{1}{N}\sum_{k=1}^N \eta_{\rm HT}'({u}_k,\lambda)\, .\nonumber
\end{eqnarray}
In this context, the term $\frac{\bm{z}^{t-1}}{\alpha} \diveta_{\rm HT} (\frac{\bm{F}^{\intercal}\bm{z}^{t-1}}{\alpha}+\bm{\hat{x}}^{t-1},\frac{\lambda}{{A}^{t-1}})$ appearing in the update of $\bm{z}^t$ is known as the Onsager reaction term, arising as a correction in the computation of the AMP iteration. In the $N\rightarrow+\infty$ limit, this extra term ensures that $\bm{z}^t$ converges to a random Gaussian vector \cite{Bayati2011,L2009}. This property allows rigorous tracking (in the $N\rightarrow +\infty$ limit) of the evolution of the mean-square error of the system. Indeed following Refs. \cite{Krzakala2012_1,Krzakala2012_2} we can derive the so-called state evolution (SE) equations
\begin{align}
\label{eq: AMP SE}
A^{t+1}&\!=\!\alpha\!\left\{\! 1\!+\!{\rm I\!E}\Big[\eta_{\rm HT}' \Big(\!\sqrt{\frac{E^t}{\alpha}}{z}\!+\!{{x}_o},\lambda/{A}^{t}\!\Big)\Big]_{{{x}_o},{z}}\!\right\}^{\!\!-1}\!\!\! , \\
m^{t+1}&\!=\!\frac{\bm{\hat{x}}^{t+1}\cdot \bxo}{N}\!=\!{\rm I\!E}\Big[ \eta_{\rm HT} \Big(\!\sqrt{\frac{E^t}{\alpha}}{z}\!+\!{{x}_o},\lambda/{A}^{t}\!\Big)x_o\Big]_{{{x}_o},{z}},\nonumber\\
q^{t+1}&\!=\!\frac{\bm{\hat{x}}^{t+1}\cdot{\bm{\hat{x}}^{t+1}}}{N}\!=\!{\rm I\!E}\Big[ \eta_{\rm HT}^2 \Big(\!\sqrt{\frac{E^t}{\alpha}}{z}\!+\!{{x}_o},\lambda/{A}^{t}\!\Big)\Big]_{{{x}_o},{z}},\nonumber
\end{align}
with the mean-squared error $E^t$ written as 
\begin{eqnarray}
E^t=\frac{\vert\vert \bxo-\bm{\hat{x}}^t\vert\vert^2_2}{N}=\E[\,x^2_o\,]_{x_o}-2m^t+q^t
\end{eqnarray}
and where ${\rm I\!E}[\cdot]_{z}$ is a Gaussian integration measure with zero mean and variance one, and ${\rm I\!E}[\cdot]_{x_o}$ is an average performed over the true signal prior from Eq.~(\ref{eq: signal distrib}).

As already detailed in a large number of publications \cite{L2009,Montanari2010, Bayati2011,Krzakala2012_1,Krzakala2012_2,Antenucci2019} and as we will see in Sec.~\ref{sec:stat mech analysis}, the AMP algorithm can probe configurations that are typical of a replica symmetric saddle-point when evaluating the expected free energy $\E \log{Z}$. We will show that to reach such configurations AMP has to be initialized with an \emph{informed} initial condition, i.e.  with $\bm{\hat{x}}^{t=0}$ already close to $\bxo$, which makes the algorithm useless in practice unless in presence of some side information on $\bxo$. 

Before analyzing the performance of AMP, we point out that a proximal gradient descent algorithm can be derived with a slight modification of Eq.~(\ref{eq: AMP algo}). Let us focus first on the iteration for $\bm{z}^t$. In fact, due to the Onsager reaction, $\bm{z}^t$ depends indirectly on the whole set of iterations $\{\bm{x}^{t'}\}_{t\geq t'\geq 0}$ and not only on the previous iteration $\bm{x}^t$. Therefore the first simplification step we can take is to approximate the iteration for $\bm{z}^t$ by changing the time indices as
\begin{eqnarray}
\label{eq: proximal gradient descent 1}
\bm{z}^t&\approx&\bm{y}-\bm{F}\bm{\hat{x}}^t +\frac{\bm{z}^{t}}{\alpha}\diveta_{\rm HT}\left(\frac{\bm{F}^{\intercal}\bm{z}^{t}}{\alpha}+\bm{\hat{x}}^{t},\frac{\lambda}{{A}^{t}}\right)\nonumber\\
\implies\!\! \bm{z}^t&\approx&\frac{\bm{y}-\bm{F}\bm{\hat{x}}^t}{1-\frac{1}{\alpha} \diveta_{\rm HT} \left(\frac{\bm{F}^{\intercal}\bm{z}^{t}}{\alpha}+\bm{\hat{x}}^{t},\frac{\lambda}{{A}^{t}}\right)}\, .
\end{eqnarray}
With this change $\bm{z}^t$ depends now only on $\bx^t$.
Then, if we replace the time dependent variable $A^t$ by a fixed value $A^\star$ we obtain
\begin{eqnarray}
\label{eq: proximal gradient descent 2}
 &&\bm{z}^t\approx\alpha\frac{\bm{y}-\bm{F}\bm{\hat{x}}^t}{A^\star}\\
&\implies&\bm{\hat{x}}^t\approx\eta_{\rm HT}\left(\frac{\bm{F}^{\intercal}\bm{z}^{t-1}}{\alpha}+\bm{\hat{x}}^{t-1},\frac{\lambda}{{A}^{\star}}\right)\nonumber\\
&\implies&\bm{\hat{x}}^t\approx\eta_{\rm HT}\left(\delta^\star\bm{F}^{\intercal}(\bm{y}-\bm{F}\bm{\hat{x}}^{t-1})+\bm{\hat{x}}^{t-1},{\lambda\delta^{\star}}\right)\, ,\nonumber
\end{eqnarray}
having defined $\delta^\star=\frac{1}{A^\star}$.
The obtained iteration for $\bm{\hat{x}}^t$ corresponds to a proximal gradient descent (PGD), with denoiser $\eta_{\rm HT}(.,.)$ and leaning rate $\delta^\star$. It can be rewritten under the equivalent form
\begin{eqnarray}
{\hat{x}}^t_i&=&\underset{x}{\rm argmin}\Big[ \frac{({\hat{x}}^{t-1}_i\!-\!x)^2}{2\delta^\star} \!-\!{F_{\mu i}}({y}^\mu\!-\!{F}^{\mu j}{\hat{x}}^{t-1}_{j})({\hat{x}}^{t-1}_i\!-\!x) +\lambda\vert\vert x\vert\vert_0\Big]
\end{eqnarray}
where we used Einstein's summation convention to obtain a compact expression.
This PGD iteration shares its fixed points with the previous AMP algorithm, granted we keep the same value for the prefactor $\lambda$. Indeed, if we consider that the AMP algorithm has a fixed point $(\bm{\hat{x}}^\star,\bm{z}^\star,A^\star)$ then Eq.~(\ref{eq: AMP algo}) imposes that
\begin{eqnarray}
\bm{\hat{x}}^\star&=&\eta_{\rm HT}\left(\frac{\bm{F}^{\intercal}(\bm{y}-\bm{F}\bm{\hat{x}}^{\star})}{A^\star}+\bm{\hat{x}}^{\star},\frac{\lambda}{{A}^{\star}}\right)\, ,\\
{{A}^{\star}}&=&\alpha- \diveta_{\rm HT} \left(\frac{\bm{F}^{\intercal}(\bm{y}-\bm{F}\bm{\hat{x}}^{\star})}{A^\star}+\bm{\hat{x}}^{\star},\frac{\lambda}{{A}^{\star}}\right)\, \nonumber
\end{eqnarray}
and
\begin{eqnarray}
\bm{z}^\star&=&\alpha\frac{\bm{y}-\bm{F}\bm{\hat{x}}^*}{A^\star}\, .
\end{eqnarray}
Therefore, comparing this result with Eqs.~(\ref{eq: proximal gradient descent 2}), we will have the same fixed point for the AMP and PGD. This property holds even when we fix $\delta^\star\neq 1/A^\star$. It can be easily proved that fixed points for $\bm{\hat{x}}^t$ with PGD (respectively with AMP) are independent of the learning rate $\delta^\star$ (respectively $A^\star$) \cite{L2009}.  

\begin{figure}
    \centering
    \includegraphics[width=0.39\textwidth]{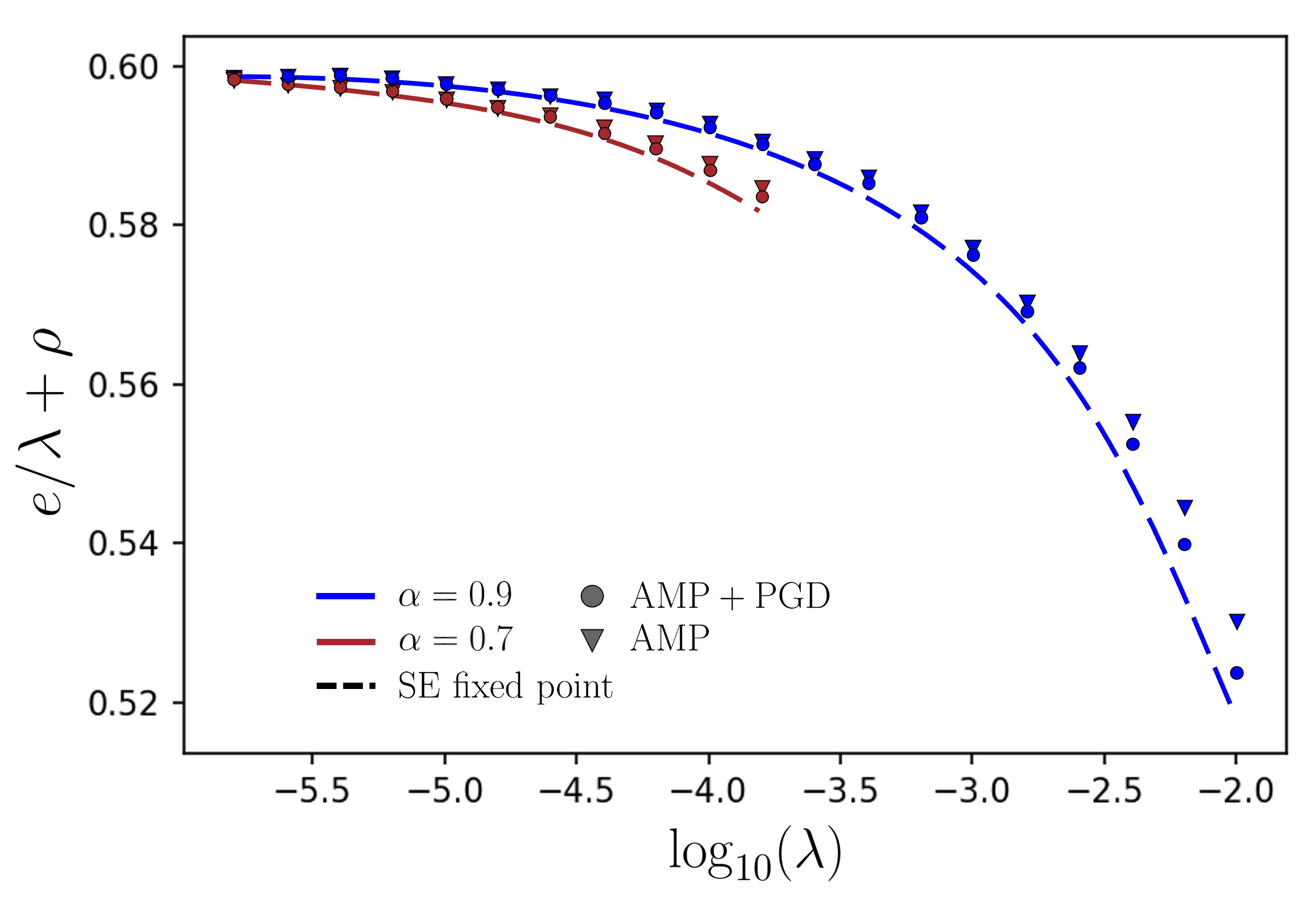}
    \includegraphics[width=0.39\textwidth]{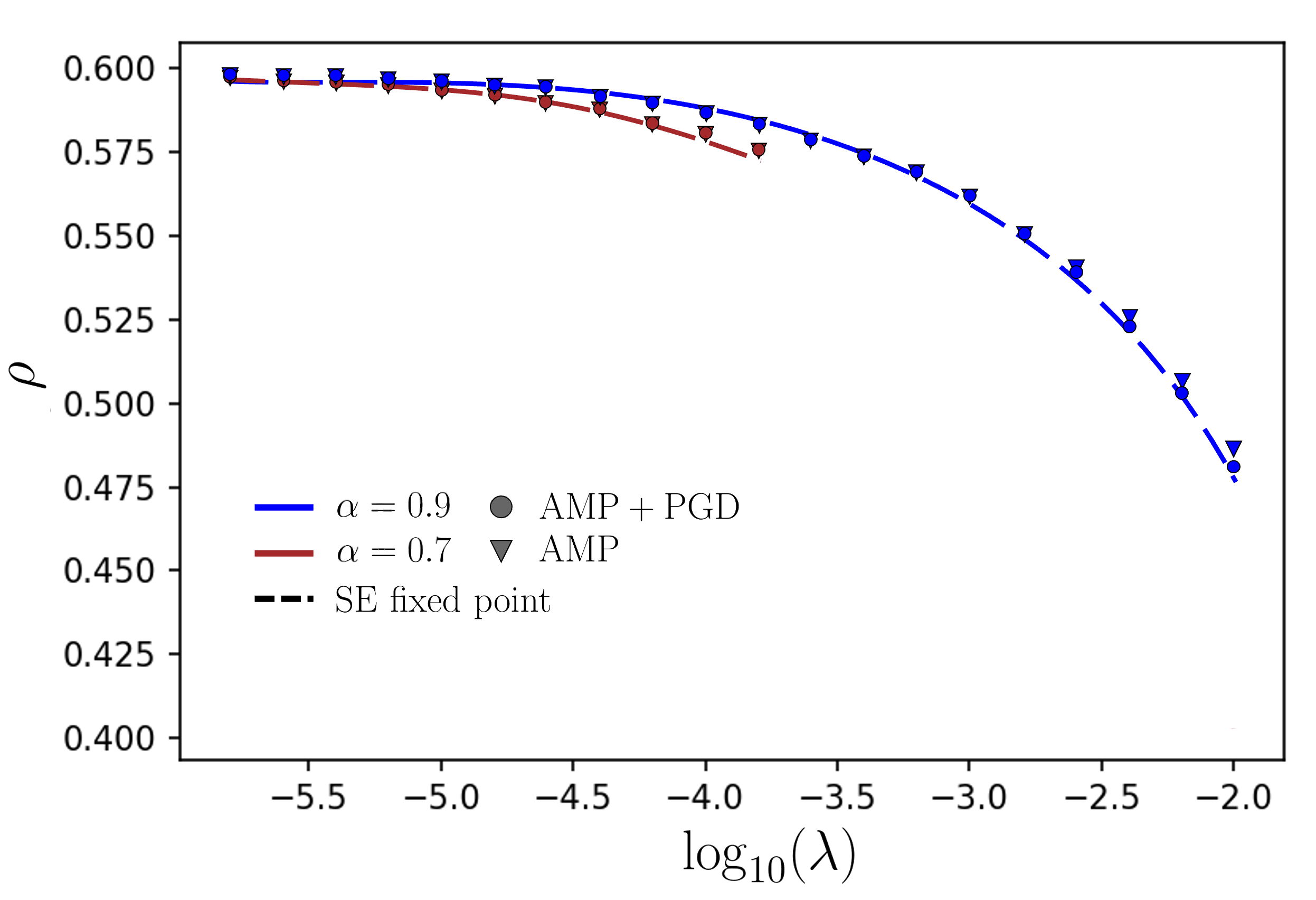}
    \includegraphics[width=0.39\textwidth]{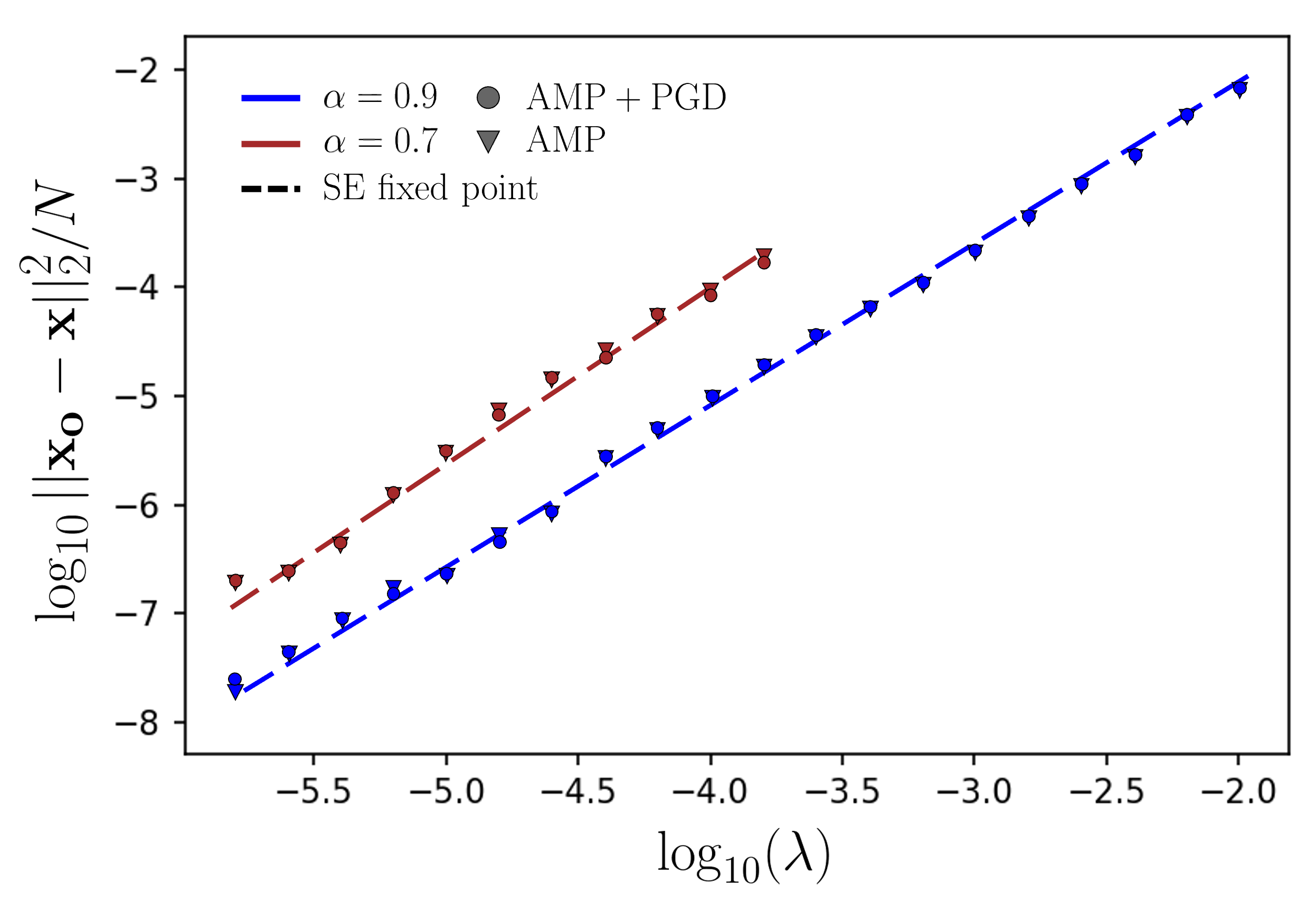} 
    \caption{Plots showing that AMP and PGD algorithms convergence near the true signal in the unrealistic setting of starting very close to it (fully-informed initialization). We simulated the AMP algorithm (triangles), and AMP followed by PGD iterations (circles) with learning rate $\delta=0.6$. In all instances, the system size is $N=5\times 10^3$ and the signal density is $\rho_o=0.6$. (\textbf{Top left}) The rescaled loss function $\left\langle\mathcal{L}_\lambda(\bx) \right\rangle /{\lambda N}=e/\lambda+\rho$ as a function of $\lambda$.  (\textbf{Top right}) Density $\rho$ of the output configuration. (\textbf{Bottom}) Reconstruction error, i.e. the $l_2$-norm distance between the signal and the output configurations. As expected the agreement between the two configurations improves when $\lambda$ is lowered. }
    \label{fig:AMP sim}
\end{figure}

In Fig.~\ref{fig:AMP sim} we show that AMP and PGD follow the predictions from the SE when an informed initialization is given. To do so, we run the AMP algorithm with initialization $\{\bm{x}^{t=0}=\bxo$, $  \bm{z}^{t=0}=0$, $A^{t=0}=\alpha\}$ and wait for a fixed point to be reached. In practice, the AMP iteration is highly unstable due to the discontinuity in the hard-thresholding denoising function. A possible cure \cite{Lucibello19a}, is to to replace the direct estimation of the variance $ \diveta_{\rm HT}(\frac{\bm{F}^{\intercal}\bm{z}^{t}}{\alpha}+\bm{\hat{x}}^{t},\frac{\lambda}{{A}^{t}})$ by  the value given by state evolution ${\rm I\!E}[\eta_{\rm HT}' (\!\sqrt{\frac{E^t}{\alpha}}{z}\!+\!{{x}_o},\lambda/{A}^{t-1}\!)]_{{{x}_o},{z}}$. When a fixed point is reached, we compute a set of observables of interest (e.g., density of non-zero entries, distance with the signal) and compare them with the state evolution predictions. As the figure shows, AMP, PGD, and state evolution predictions are found to be in good agreement.

In Fig.~\ref{fig:AMP sim}, we also check that running a proximal gradient descent starting from an AMP fixed point does not modify drastically the configuration yielded by the algorithm. 
Note that, for stability reasons, we run PGD with learning rate $\delta=0.6$, smaller than the value $\delta^\star=1/A^\star$ obtained with the AMP fixed point.
As can be seen from the top panels, when the PGD is run after the AMP iterations, we observe a better agreement with the predictions from the state evolution fixed point (dashed lines) both for the rescaled loss function and for the density of the output configuration. On the bottom, we instead see that the extra PGD steps do not lead to an observable improvement of the reconstruction error. These observations indicate that PGD simply sets the small magnitude entries of the AMP fixed point to zero.
Finally, in the bottom panel we also observe that when $\alpha>\rho_o$, the recovery of the signal improves as the regularization is sent to zero ($\lambda\rightarrow0$). 

Note that, both the AMP and the derived PGD require an initialization around the signal $\bxo$. If instead a random initialization for $\bm{\hat{x}}^{t=0}$,  $\bm{z}^{t=0}$ and $A^{t=0}$ is considered, these algorithms will simply diverge. This last well-known feature makes these hard-thresholding denoising methods useless for true agnostic signal recovery.

\subsection{Approximate Survey Propagation for $\lzero$ reconstruction}
\label{subsec: ASP and ASPo}§

Having discussed the failure of AMP and PGD algorithms in recovering the signal when starting from an uninformed initialization, we will now see how Approximate Survey Propagation (ASP) instead will succeed with a compression rate close to the optimal one. The ASP algorithm has been introduced in Ref. \cite{Antenucci2019} for pairwise systems and in Ref. \cite{Lucibello19a} in the context of generalized linear models. 
It is closely related to the 1-step Replica Symmetry Breaking formalizing discussed in Section \ref{sec:stat mech analysis}.
It can be derived from the AMP framework when applied to a systems containing $s$ real replicas, $s$ being the Parisi-parameter to be taken as integer first and the analytically continued to real values in the message passing equations. 
Following Refs. \cite{Antenucci2019,Lucibello19a} (see also Ref. \cite{chen2023vasp} for a generalization beyond Gaussian matrices), we obtain the message passing equations for finite $\beta$ and $\varepsilon$. Then, taking the limited $\beta\to+\infty$ followed by $\varepsilon\to 0^+$ we end up with the final form of the algorithm below, with $s$ and $\lambda$ left as hyper-parameters.
Given the initialization $\bm{\hat{x}}^{t=0}=0$, ${V}_1^{t=0}={V}_0^{t=0}=0$, $\bm{g}^{t=0}=0$ the ASP iteration reads
\begin{align}
    \label{eq: ASP algo}
    w_\mu^t&=\sum_i F_{\mu i}\hat{x}_i^{t-1}-{g}_\mu^{t-1}({s}\, {V}_0^{t-1}+ {V}_1^{t-1}) \, ,\\
     {g}_\mu^t&=\frac{y_\mu-w_\mu^t}{1+{V}_1^{t-1}+s{V}_0^{t-1}},\nonumber\\
    {A}_0^{t}&=\frac{\alpha}{s}\left(\frac{1}{1+V_0^{t-1}}-\frac{1}{1+{V}_1^{t-1}+s{V}_0^{t-1}}\right) \nonumber\, ,\\
    {A}_1^{t}&=\frac{\alpha}{1+{V}_1^{t-1}+s{V}_0^{t-1}}+s{A}_0^{t}
    \nonumber\, ,\\
    {B}_i^{t}&=\sum_\mu F_{\mu 
    i}{g}_\mu^t+\hat{x}_i^{t-1}({A}_1^{t}-{s}{A}_0^{t}) \nonumber\, ,\\
    \hat{x}_i^{t} &=\partial_{{B}} \phi^{\rm 1RSB}_{in}({B}_i^t,{A}_0^t,{A}_1^t) ,\nonumber\\
    {V}_0^{t}&=\frac{\sum_i\left[-2\partial^2_{{A}_1}\phi^{\rm 1RSB}_{in}({B}_i^t,{A}_0^t,{A}_1^t)+(\hat{x}_i^t)^2\right]}{N} \nonumber\, ,\\
    {V}_1^{t}&=\partial^2_{{B}} \phi^{\rm 1RSB}_{in}({B}_i^t,{A}_0^t,{A}_1^t)-{s}\,{V}_0^{t}, \nonumber\, 
\end{align}
with
\begin{align}
\phi^{\rm 1RSB}_{in}(B,A_1,A_0)&=\frac{1}{s}\log\E\left[  
\,e^{s\max\left(\frac{(B+\sqrt{A_0}z)^2}{2A_1}-\lambda,0\right) }
\right]_z\!.
\end{align}
Here ${\rm I\!E}[\cdot]_{z}$ is expectation over a standard Gaussian random variable. As we will see later in Sec.~\ref{sec:stat mech analysis}, $\phi^{\rm 1RSB}_{in}$ corresponds to the entropic contribution when estimating the free energy within a 1RSB ansatz. In the zero-temperature limit, the role of the parameter $s$ is to select a family of states among the exponentially many available, tuning the balance between the energy of the state $e + \lambda \rho$ and their log-number (complexity) $\Sigma(e +\lambda \rho)$. The states with the lowest energy are the less numerous ones and can be selected by choosing an $s$ such that $\Sigma = 0$. 
Therefore during the message passing iterations Ref. \ref{eq: ASP algo}, we adaptively select $s$ in order to satisfy at each time the $\Sigma=0$ condition, which reads 
\begin{eqnarray}
\label{eq: ASP zero complexity}
    \frac{A_0^t}{2}\left(\frac{1}{N}\sum_i^N (\hat{x}_i^t)^2+V_0^t\right)\!-\frac{\alpha}{2s}\Bigg[\frac{1}{s}\ln\Big(\frac{1+V_1^t}{1+V_1^t+sV_0^t}\Big)+\frac{V_0^t}{1\!+\!V_1^t\!+\!sV_0^t}\Bigg]\!+\frac{1}{N}\!\sum_i\partial_s \phi^{\rm 1RSB}_{in}({B}_i^t,{A}_0^t,{A}_1^t)\!=0\,.
\end{eqnarray}
The ASP algorithm, even in the adaptive $s$ formulation, can be tracked by the low dimensional set of SE equations \cite{Antenucci2019}:
\begin{align}
\label{eq: ASP SE}
    A^{t+1}_1\!-\!sA_0^{t+1}&\!=\!\alpha\!\left\{\! 1\!+\!{\rm I\!E}\Big[\partial^2_{{B}} \phi^{\rm 1RSB}_{in} \Big(\sqrt{\frac{E^{t}(A_1^t\!-\!sA_0^t)^2}{\alpha}}{z}+(A_1^t-sA_0^t){{x}_o},{A}_0^{t},{A}_1^{t}\Big)\Big]_{{{x}_o},{z}}\!\right\}^{-1} , \\
    A^{t+1}_1&\!=\alpha\Bigg\{ 1\!+\!q^{t}-\!2{\rm I\!E}\Big[\partial^2_{{A_1}} \phi^{\rm 1RSB}_{in} \Big(\sqrt{\frac{E^{t}(A_1^t\!-\!sA_0^t)^2}{\alpha}}{z}\!+\!(A_1^t\!-\!sA_0^t){{x}_o},{A}_0^{t-1},{A}_1^{t-1}\Big)\Big]_{{{x}_o},{z}}\!\Bigg\}^{-1} , \nonumber\\
    m^{t+1}&={\rm I\!E}\Big[x_o\partial_{{B}} \phi^{\rm 1RSB}_{in} \Big(\sqrt{\frac{E^{t}(A_1^t\!-\!sA_0^t)^2}{\alpha}}{z}\!+(A_1^t\!-\!sA_0^t){{x}_o},{A}_0^{t},{A}_1^{t}\Big)\Big]_{{{x}_o},{z}} , \nonumber\\
    q^{t+1}&={\rm I\!E}\Bigg\{\Big[\partial_{{B}} \phi^{\rm 1RSB}_{in} \Big(\sqrt{\frac{E^{t}(A_1^t\!-\!sA_0^t)^2}{\alpha}}{z}\!+(A_1^t\!-\!sA_0^t){{x}_o},{A}_0^{t},{A}_1^{t}\Big)\Big]^2\Bigg\}_{{{x}_o},{z}} . \nonumber
\end{align}

Inspection of the behavior of the ASP algorithm in the limit of zero regularization intensity suggests further simplifications in the iteration. Indeed, we observe that zero-complexity $s$ diverges to infinity as $\lambda$ goes to $0^+$, with $s A_0$ and $s V_0$ remaining finite. 
This behavior inspires the proposal of a simplified message-passing algorithm that we name ${\rm ASP_o}$. Given the initialization $\bm{\hat{x}}^{t=0}=0,\bm{{z}}^{t=0}=0,A^{t=0}=\alpha$, the ${\rm ASP_o}$ iterations  for $t\geq 1$ read:
\begin{align}
\hat{x}_i^t&\!=\!\eta_{\rm ASP_o}\!\left(\frac{B_i^t}{A^t}\!=\!\sum_{\mu=1}^M {F}_{\mu i}z_\mu^{t-1}+\hat{x}_i^{t-1},\frac{\lambda}{A^t},\xi\right) , \\
z_\mu^t&=y_\mu\!-\!\sum_{i=1}^N {F}_{\mu i}x_i^{t-1}\!+\!\frac{z_\mu^{t-1}}{\alpha }{V}^{t-1},\nonumber \\
{A}^{t}&\!=\!\frac{\alpha}{1\!+\!{V}^{t-1}}\;\, \text{,}\; \,\, {V}^{t}=\sum_{k=1}^N \partial_{u} \eta_{\rm ASP_o}\!\left(\!u\!=\!\frac{B_i^t}{A^t},\frac{\lambda}{A^t},\xi\right),\; \,\nonumber
\end{align}
with denoising function 
\begin{equation}
\eta_{\rm 
ASP_o}\!\!\left(u,\lambda,\xi\right)\!=\!u\Big[1\!-\!\frac{1}{2}\erfc\Big(\frac{u\!-\!\sqrt{2\lambda}}{\xi\lambda}\Big)\!+\!\frac{1}{2}\erfc\Big(\frac{u\!+\!\sqrt{2\lambda}}{\xi\lambda}\Big)\Big]. 
\end{equation}
In particular we replaced the variables $V_0^t$ and $V_1^t$ (respectively $A_0^t$ and $A_1^t$) with the combination $V^t=V_1^t+sV_0^t$ (respectively $A^t=A_1^t-sA_0^t$). As it is clear from its structure, also ${\rm ASP_o}$ belongs to the AMP family. We also highlight the introduction of a new parameter $\xi$, which corresponds in practice to a ``smoothing" parameter in the denoiser. To understand this ``smoothing" we can shortly consider the two limiting cases. First, when $\xi=0$ the function $(1/2) \erfc\cdot/\xi$) boils down to a Heaviside function $\Theta(\cdot)$ and we retrieve the hard-thresholding denoiser,  $\eta_\mASPo(u\,,\lambda\,,\xi=0)=u\, \Theta(\vert u\vert-\sqrt{2\lambda})$, that we expect when introducing a $\lzero$ norm penalty. Secondly, if we take $\xi=+\infty$ the denoiser becomes the identity function: $\eta_\mASPo(u\,,\lambda\,,\xi=+\infty)=u$. Thus any intermediate value of $\xi$ gives rise to a smoothed version of the hard-thresholding denoiser.

Both algorithms, ASP and ${\rm ASP_o}$, start from uninformed initialization ($\bm{\hat{x}}^{t=0}=0$). We observe that, for sufficiently large $\alpha$, switching off adiabatically the regularizer prefactor $\lambda$ leads to signal recovery. In particular ${\rm ASP_o}$ appears to be the most efficient algorithm of the two, as it does not require any additional tuning for fixing $\Sigma=0$. Indeed, it can be noted that the parameter $s$ has disappeared in ${\rm ASP_o}$. Besides, setting a value for $\xi$ is an easy task, starting the algorithm at a given value $\lambda$ we simply choose $\xi$ big enough for the algorithm to be stable and to converge to a fixed point. Then, $\lambda$ can be adiabatically set to zero without having to adjust $\xi$ again.

\begin{figure}[t]
    \centering
    \includegraphics[width=0.49\textwidth]{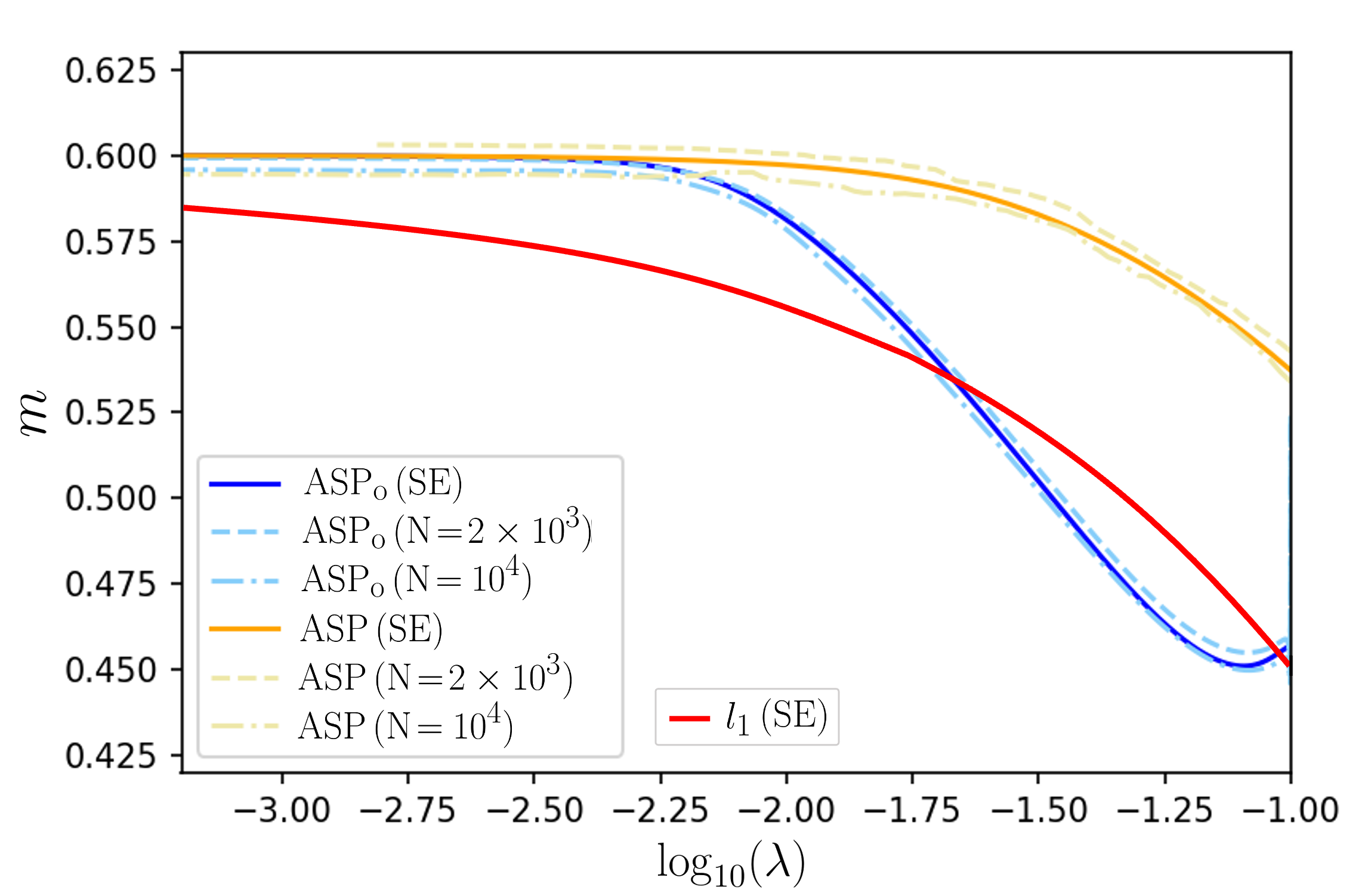}
    \includegraphics[width=0.49\textwidth]{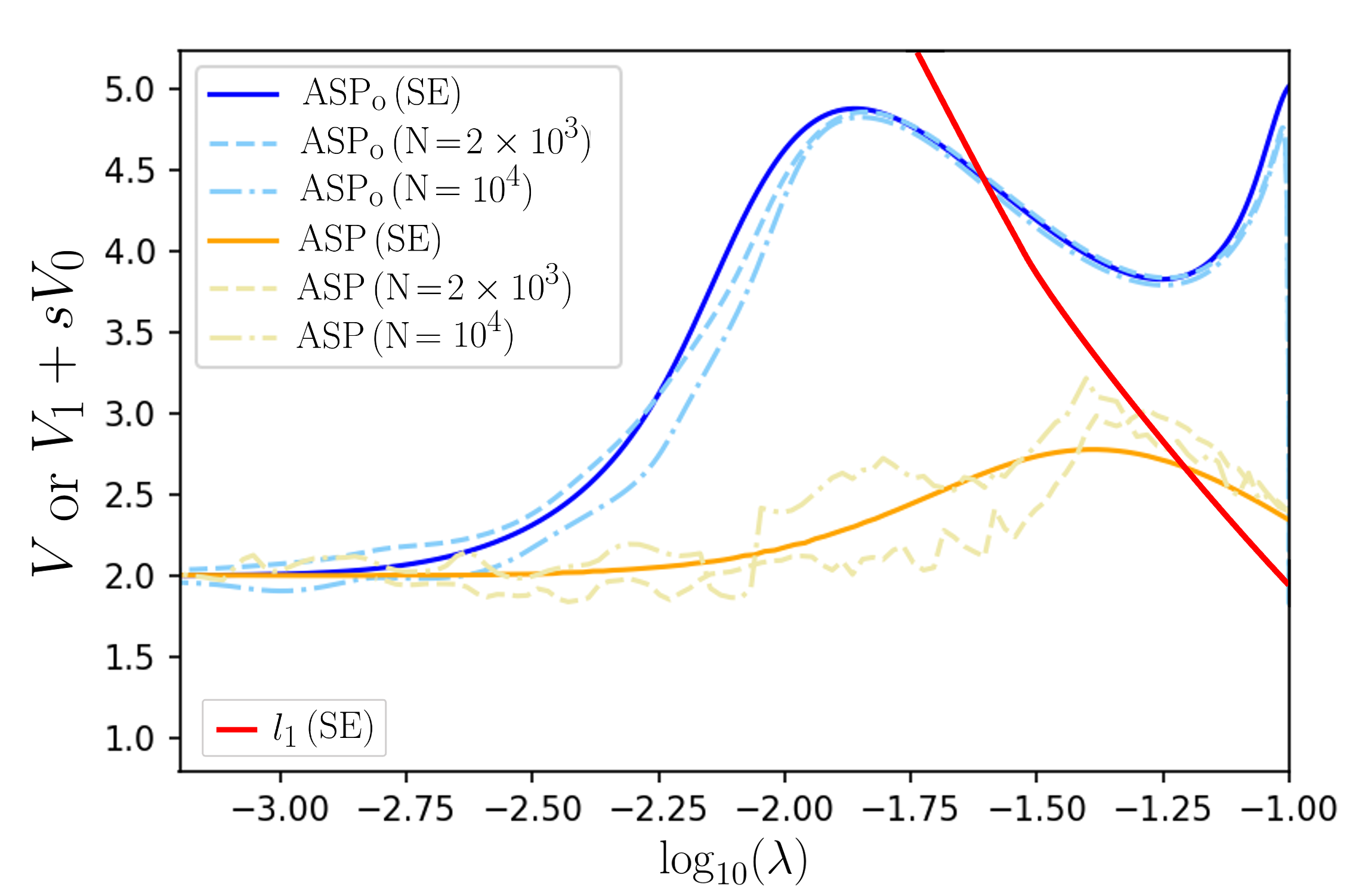}
    \caption{(\textbf{Left}) Overlap at convergence with the true signal when running the ASP (orange), ${\rm ASP_o}$ (blue) and LASSO (red) algorithms at different values of the regularization prefactor $\lambda$. The dashed lines correspond to finite size simulations while the full lines correspond to the infinite size prediction from State Evolution (SE). For this experiment, we set $\rhoo=0.6$ and $\alpha=0.87$. The ${\rm ASP_o}$ parameter is fixed to $\xi=0.7$.  (\textbf{Right}) The quantity $V$ (for ${\rm ASP_o}$ and LASSO) and $V_1+s V_0$ (ASP) in the same experimental setting.}
    \label{fig: ASP ASPo signal recovery}
\end{figure}
In Fig.~\ref{fig: ASP ASPo signal recovery}, we show the behavior of ASP and ${\rm ASP_o}$ in a regime where they achieve signal recovery (as predicted by the 1RSB free energy) and where the LASSO algorithm instead fails at this task. Each observable is computed once the algorithmic iterations converge to a fixed point. First, we can note that the magnetization $m=\bxo\cdot \bhx^{t=+\infty}/{N}$  converges to $\rhoo$ as $\lambda$ is sent to $0^+$, implying perfect signal reconstruction. As mentioned earlier, ${\rm ASP_o}$ corresponds to a low $\lambda$ limit of ASP, which explains why ASP and ${\rm ASP_o}$ output not only the same value for $m$ but e.g. also the same value for $V= V_1+sV_0$ when $\lambda$ goes to zero. In parallel, we rigorously tracked the performances of our algorithms in the $N\rightarrow + \infty$ limit via state evolution (SE) \cite{L2009,Antenucci2019}; we plotted these predictions in full lines. In the case of ${\rm ASP_o}$, the state evolution consists in the set of equations
\begin{align}
\label{eq: ASPo SE}
A^{t+1}&\!=\!\alpha\!\left\{\! 1\!+\!{\rm I\!E}\Big[\eta_\mASPo' \Big(\!\sqrt{\frac{E^t}{\alpha}}{z}\!+\!{{x}_o},\lambda/{A}^{t}\!\Big)\Big]_{{{x}_o},{z}}\!\right\}^{-1}\!\! , \;\;\;\;\;\nonumber\\
    m^{t+1}&= \E\!\left[x_o\,\eta_\mASPo\!\left(\frac{\sqrt{E^t}z}{A^t}+x_o,\frac{\lambda}{A^t},\xi\right)\right]_{x_o,z}\, ,\\
    q^{t+1}&= \E\!\left[ \eta^2_\mASPo\!\left(\frac{\sqrt{E^t}z}{A^t}+x_o,\frac{\lambda}{A^t},\xi\right)\right]_{x_o,z}\, .\nonumber
\end{align}
Finally, in Fig.~\ref{fig: ASP ASPo signal recovery} we also show the outputs for $m$ and $V$ given by the SE equations with the LASSO denoising function (also called $l_1$-based AMP). In this regime, the LASSO algorithm is not able to recover the signal.

  \begin{figure}[t]
      \centering      \includegraphics[width=0.49\textwidth]{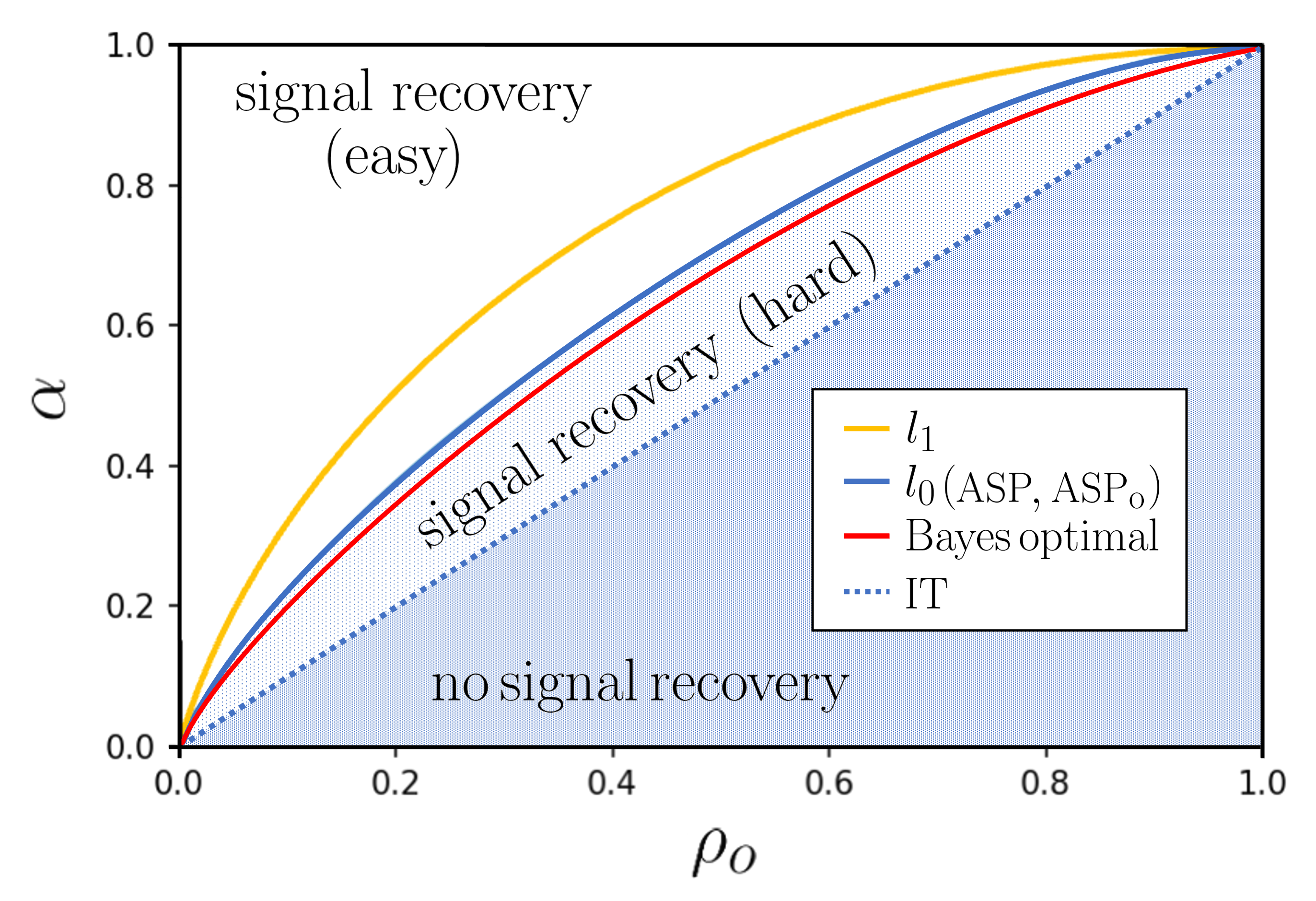}
    \includegraphics[width=0.49 \textwidth]{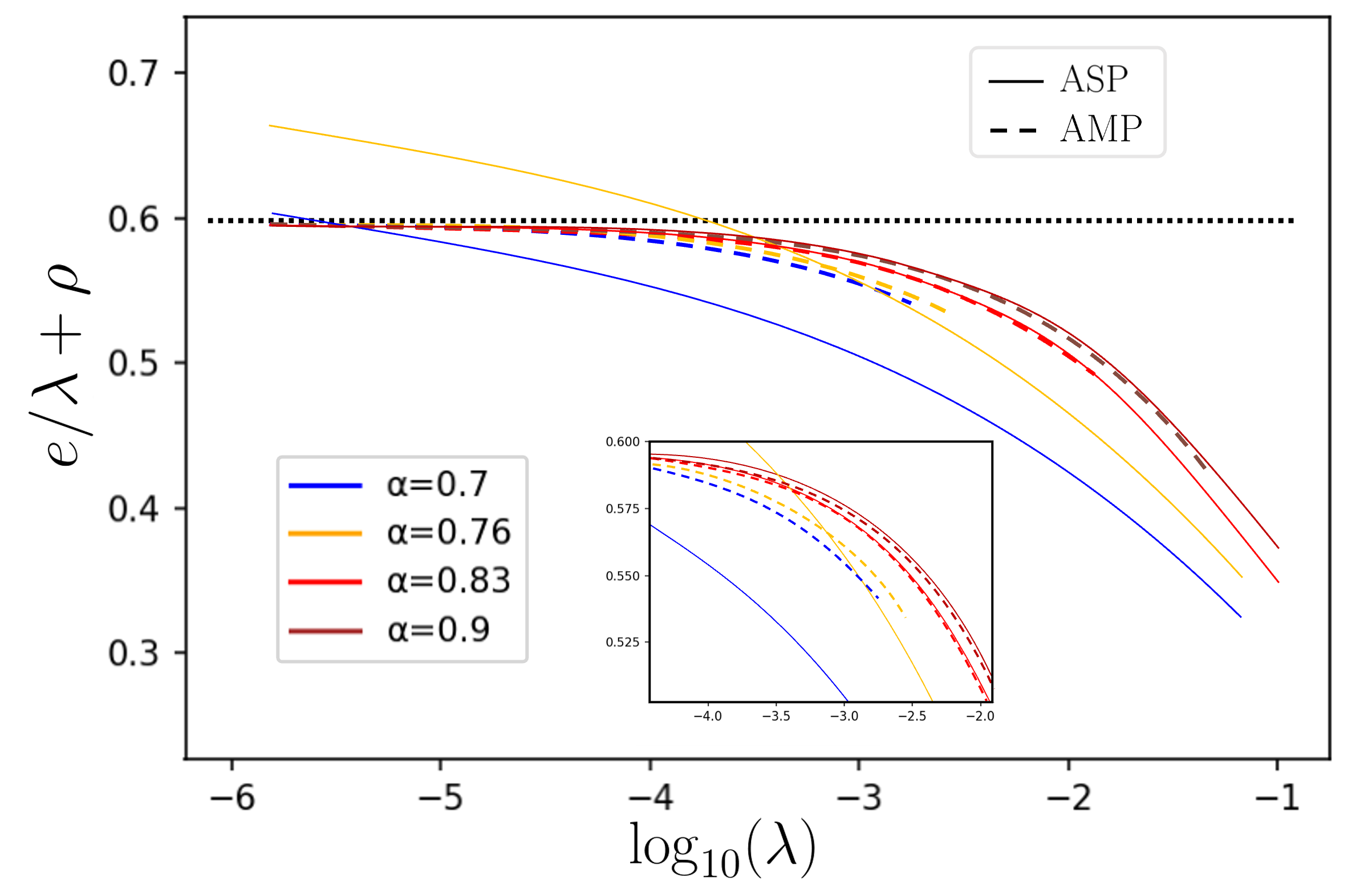}
      \caption{(\textbf{Left}) Phase diagram for perfect recovery in compressed sensing. The $\lone$ and Bayesian Optimal lines give the algorithmic thresholds for $\lone$ based reconstruction and Bayesian optimal AMP respectively. Above those lines, the corresponding algorithms succeed with high probability for large systems. The $\lzero$ (ASP, ${\rm ASP_o}$) line instead, gives the recovery threshold for our algorithms as predicted by state evolution. Lastly, the  IT line is the information theoretic threshold for perfect recovery. It also corresponds to the ultimate limit for $\lzero$-based recovery algorithms. In the region between  the IT and the $\lzero$ (ASP, ${\rm ASP_o}$) line though, it is algorithmically hard to find the global optimum of the $\lzero$-based loss.
      (\textbf{Right}) We plot  the rescaled cost $\left\langle\mathcal{L}_\lambda(\bx)\right\rangle /{\lambda N}=e/\lambda+\rho$ as a function of $\lambda$, for $\rhoo=0.6$ and several values of $\alpha$. We show the predictions from the informed AMP state evolution (i.e. $m^{t=0}\approx q^{t=0}\approx\rho_o$) and the uninformed ASP one. In the regime for which the signal recovery is easy ($\alpha>0.83$) both saddle points describe the true signal as $\lambda$ goes to zero, we thus have $e/\lambda+\rho\to \rhoo$.
      At lower $\alpha$ instead, the uninformed saddle point no longer describes the true signal, but configurations at larger density.}
      \label{fig: 1RSB computation}
  \end{figure}

 Analyzing the non-trivial fixed point for the ASP state evolution equations makes it possible to determine for which range of parameters $(\alpha,\rho_o)$ our two algorithms are capable of perfect signal recovery. Indeed, since ${\rm ASP_o}$ follows from the ASP, the two algorithms have the same performances regarding signal recovery. As shown in the right panel of Fig.~\ref{fig: 1RSB computation}, if $\alpha$ is sufficiently large ($\alpha\geq0.83$ for $\rho_o=0.6$) we observe that adiabatically switching off the $l_0$ norm regularization with ASP yields $e/\lambda+\rho\rightarrow\rho_o$. This is indicative, with state evolution, of perfect signal recovery. For smaller values of $\alpha$, survey propagation instead returns states with a larger cost function compared to that of the signal. The transition between these two regimes draws the limit for which ASP and ${\rm ASP_o}$ are capable of perfect signal recovery. As expected the hard-thresholding AMP with the informed initialization always returns the signal as long as $\alpha>\rho_o$, as it yields $e/\lambda+\rho\rightarrow\rho_o$. In the left panel of Fig.~\ref{fig: 1RSB computation}, we display the transition lines for perfect signal recovery using LASSO, Bayes optimal AMP, and the two survey-based algorithms. ASP and ${\rm ASP_o}$ outperform LASSO, the standard agnostic algorithm for compressed sensing. Understandably, the Bayes optimal setting still offers a performance improvement, but the knowledge of the signal distribution is necessary for this algorithm to work.

\section{Statistical physics analysis}
\label{sec:stat mech analysis}
In this section, we focus on computing the asymptotic average free energy associated with this problem, i.e. the expectation of the logarithm of the partition function, from Eq.~\eqref{eq: partition function}, in the large $N$ limit. 
As already pointed out in numerous publications \cite{Antenucci2019,L2009,Montanari2010,Krzakala2012_1,Krzakala2012_2,Lucibello19a,Bereyhi2017,Bereyhi2018,Bereyhi2019,Kabashima2009}, there is a direct connection between the algorithms presented above (in particular hard thresholding AMP and ASP) and the computation of this free energy via replica method. We estimate the average of $\log{Z}(\by,\bF,\lambda)$ via the identity
\begin{align}
\E\left[\log{Z}(\by,\bF,\lambda)\right]_{\bxo,\bF} \underset{n\rightarrow 0}{=}\frac{\E\left[ {Z^n}(\by,\bF,\lambda)\right]_{\bxo,\bF}\,-\,1}{n}\, , 
\end{align}
where the expectation over $\by$ is 
implicitly entailed in the expectation over $\bxo$ and $\bF$. 
The right-hand side of the previous equation is first computed for $n\in \N$ and the result is then extended for $n \in \R^+$. Each of the $n$ copies of the system implied in ${Z^n}(\by,\bF,\lambda)$ is called a replica, and has a corresponding configuration labeled as $\bx^{a}$ ($a=1,\dots,n$). We compute $ \E\left[{Z^n}(\by,\bF,\lambda)\right]_{\bxo,\bF}$ through a saddle-point approximation, where the action is a function of the overlap matrix between replicas:
\begin{align}
Q^{ab}=\frac{1}{N}\langle \bx^{a}\cdot \bx^{b}\rangle\,  \;\quad a,b=1,\dots,n.
\end{align}
Here $\cdot$ denotes scalar product and again $\langle \cdot \rangle$ indicates the average over the distribution $P(\bm{x})$. In the following, the computation will be performed under the assumption of replica symmetry (RS) or at one step of replica symmetry breaking (1RSB) \cite{Mezard1987,Mezard09}. In particular, we will outline the connection between the RS (respectively 1RSB) computation and the AMP (respectively ASP) algorithm.

\subsection{RS $Ansatz$}
\label{subsec: rs ansatz}
Following previous work on compressed sensing \cite{Kabashima2009,Kabashima2010,Krzakala2012_1,Krzakala2012_2} we will focus in this section on the replica symmetric (RS) $Ansatz$:
\begin{align}
Q^{ab}=\left\{
    \begin{array}{ll}
       Q & \text{if}\; a= b \\
       q & \text{if}\; a\neq b \\
    \end{array}
\right.\, .
\end{align}

For the saddle-point approximation to be self-consistent we also have to introduce the magnetization of the replicas with respect to the signal,
\begin{eqnarray}
m=\frac{1}{N}\langle\bm{x}^a \cdot \bxo  \rangle\, , \quad a=1,\dots , n\, .
\end{eqnarray}
The finite temperature free energy is detailed in App.~\ref{app: finite temp free energy}. In the following we will focus on the zero-temperature limit which reads
\begin{eqnarray}
\label{eq: RS zero-temp free energy}
\frac{1}{\beta N}{\rm I\!E}\left[\log[{Z}(\bm{y},\bm{F},\lambda)]\right]_{\bm{y},\bm{F}}&=&\underset{V,A,q,\hat{q},m,\hat{m}}{\rm extr}\Bigg\{\!\!-\!\hat{m}m+\frac{Aq-\hat{q}V}{2}-\frac{\alpha}{2}\frac{{\rm I\!E}[{x_o}^2]_{{x_o}}-2m+q}{1+V}\\
&&\hspace{2cm}+{\rm I\!E}\Bigg[ {\rm max}\left\{\frac{(z\sqrt{\hat{q}}+\hat{m}x_o)^2}{2A}-\lambda,0\right\}\Bigg]_{x_o,z}\;\Bigg\}\nonumber\;.
\end{eqnarray}
where, again, we used the short-hand notation $\E[\cdot]_z$ to indicate an expectation over a scalar normal-distributed variable $z$, and ${\rm I\!E}[\cdot]_{x_o}$ indicates the average over one entry of the signal, from Eq.~(\ref{eq: signal distrib}).
Note that to have a meaningful replica $Ansatz$ we need $Q>q$ and consequently $V>0$, as $V=\beta(Q-q)$.

The connection between the RS computation and the AMP becomes clear when comparing the free energy saddle-point equations with the SE from Eq.~\ref{eq: AMP SE}. Indeed, the saddle-point equations are
\begin{align}
\label{eq: RS saddle point}
    \hat{q}&=\alpha\frac{{\rm I\!E}[{x_o}^2]_{{x_o}}-2m+q}{(1+V)^2},\\
    \hat{m}&=A\nonumber\\
     V&=1\!+\!{\rm I\!E}\left[\eta_{\rm HT}' \Big(\!\sqrt{\frac{\hat{q}}{A^2}}{z}\!+\!{{x}_o},\lambda/{A}\!\Big)\right]_{{{x}_o},{z}},\nonumber\\
     A&=\frac{\alpha}{1+V},\nonumber \\
m&={\rm I\!E}\left[ \eta_{\rm HT} \Big(\!\sqrt{\frac{\hat{q}}{A^2}}{z}\!+\!{{x}_o},\lambda/{A}^{t}\!\Big)x_o\right]_{{{x}_o},{z}},\nonumber\\
q&={\rm I\!E}\left[ \eta_{\rm HT}^2 \Big(\!\sqrt{\frac{\hat{q}}{A^2}}{z}\!+\!{{x}_o},\lambda/A\!\Big)\right]_{{{x}_o},{z}}.\nonumber
\end{align}
This set of equations is nothing but the AMP state evolution equations evaluated at a fixed point, i.e. with all quantities being stationary and time-independent.  

For $\alpha>\rho_o$, we recall that the RS $Ansatz$ admits one stable saddle-point for sufficiently low values of the regularization prefactor $\lambda$. This solution converges to the signal $\bxo$ for $\lambda\rightarrow 0^+$ as this saddle-point verifies $\underset{\lambda\rightarrow 0^+}{\lim}{\rm I\!E}\left[\vert\vert\bxo- \bm{x}^a\vert\vert_2^2\right]_{\bxo,\bm{F}}=0$. With the RS computation a simple argument enables to justify that such a solution can only be obtained for $\alpha>\rho_o$. When considering the saddle-point Eqs.~(\ref{eq: RS saddle point}), we have with signal retrieval
\begin{eqnarray}
A=\frac{\alpha}{1+V}>0\; ,\quad V\underset{\lambda\rightarrow 0^+}{=}\frac{\rho(\lambda)}{A}>0
\end{eqnarray}
with the estimated density of the signal
\begin{eqnarray}
\rho(\lambda)\!=\!{\rm I\!E}\Bigg[{\rm max}\!\left\{\frac{(z\sqrt{\hat{q}}+\hat{m}x_o)^2}{2A}-\lambda,0\!\right\}\Bigg]_{{x_o}, z}\!.
\end{eqnarray}
Closing the equation for $V$ we have
\begin{eqnarray}
V\underset{\lambda\rightarrow 0^+}{=}\frac{\rho(\lambda)}{\alpha-\rho(\lambda)}>0\, ,
\end{eqnarray}
therefore to ensure both $\rho(0^+)=\rho_o$ and $V>0$ we have to be in a regime where $\alpha>\rho_o$. In other words, this means that the transition for signal recovery predicted by the RS computation exactly matches the information theoretic threshold. 


Finally, the ty of the RS saddle-point has often been discussed \cite{Kabashima2009,Kabashima2010} using the Almeida-Thouless (AT) stability criteria \cite{Almeida_1978}
\begin{align}
    \frac{\alpha{\rm I\!E}\left[\partial^2_B{\rm max}\!\left.\left\{\frac{B^2}{2A}-\lambda,0\!\right\}\right\vert_{B=z\sqrt{\hat{q}}+\hat{m}x_o}\right]_{{x_o}, z}}{(1+V)^2}<1\, .
\end{align}
However, it can be rapidly checked that the term involves an ill-defined integral over $x_o$ and $z$. Indeed, the double derivative over $B$ yields a squared Dirac distribution and thus we have
\begin{align}
    {\rm I\!E}\left[\partial^2_B{\rm max}\!\left.\left\{\frac{B^2}{2A}-\lambda,0\!\right\}\right\vert_{B=z\sqrt{\hat{q}}+\hat{m}x_o}\right]_{{x_o}, z}=+\infty\,.
\end{align}
Due to this result, it has often been concluded that the RS saddle point is unstable, as it does not satisfy the AT stability criteria. Nevertheless, this criterion assumes that the free energy can be Taylor expanded in the overlap space $\{Q^{ab}\}_{a,b\in [\![1,n]\!]}$ around the RS $Ansatz$. In Sec.~\ref{sec: RS stability} we show, using the 1RSB free energy, that this assumption is incorrect in this setting. Moreover, we show via a more subtle analysis that the RS saddle point is in fact stable.

\subsection{1RSB $Ansatz$}
\label{subsec: 1-rsb ansatz}
In this section, we will consider the more general one-step replica symmetry breaking (1RSB) \ansatz  \cite{mezard1987spin}. According to the 1RSB prescription, the $n$ replicas are arranged in blocks of equal size $s\beta^{-1}$, and the $Q$ matrix is parametrized as follows:
\begin{align}
Q^{ab}=\left\{
    \begin{array}{ll}
       q_0+ \beta^{-1}V_1 +V_0 & \text{if}\; a= b \\
       q_0+V_0 & \lfloor\frac{a}{s\beta^{-1}}\rfloor= \lfloor\frac{b}{s\beta^{-1}}\rfloor \\
       q_0 & \text{otherwise}\\
    \end{array}
\right.\! .
\end{align}
Under these assumptions and in the $\beta\to\infty$ limit, the free energy reads \cite{Antenucci2019,Obuchi2018,Bereyhi2017} 
\begin{align}
    \lim_{   \beta\to \infty }    \lim_{   N\to \infty }
    \frac{1}{\beta N}\E \log{Z}(\bxo,\bF,\lambda)=\sup_{s\geq 0}  \,\Phi(s),
\end{align}
with the free entropy $\Phi$ defined as
\begin{align}
\label{eq:free_energy_1rsb}
&\Phi(s)=\extr\Bigg\{-\hat{m}m+\frac{(V_0+q_0)A_1-V_1(\hat{q}_0+A_0)}{2}+\frac{{s}(q_0+V_0)(\hat{q}_0+A_0)+sq_0\hat{q}_0}{2}\\
&\hspace{2cm}+\E\Big[ {\phi}^{\rm 1RSB}_{in}(z\sqrt{\hat{q_0}}+\hat{m}x_o,A_1,A_0,\lambda)\Big]_{{x_o}, z} -\frac{\alpha}{2}\frac{\E[{x_o}^2]_{{x_o}}-2m+q_0}{1+V_1+{s}V_0}+\frac{\alpha}{2{s}}\log\Big(\frac{1+V_1}{1+V_1+{s}V_0}\Big)\Bigg\},\nonumber
\end{align}
and where again we used the short-hand notation $\E[\cdot]_z$ to indicate an expectation over a scalar normal-distributed variable $z$. The finite temperature free energy is detailed in App.~\ref{app: finite temp free energy}. In Eq. \eqref{eq:free_energy_1rsb}, extremization is performed with respect to the parameters $ V_0,A_0,V_1,A_1, q_0,\hat{q}_0,m,$ $\hat{m}$,
and $\phi^{\rm 1RSB}_{in}$ is defined as 
\begin{align}
\label{eq:free_energy_1rsb in channel}
\phi^{\rm 1RSB}_{in}(B,A_1,A_0,\lambda)&=\frac{1}{s}\log\E\left[  
\,e^{s\max\left(\frac{(B+\sqrt{A_0}z)^2}{2A_1}-\lambda,0\right) }
\right]_z\!.
\end{align}
The parameter $s$ allows probing of the 
different energy levels for $\mathcal{L}$. 
Within the 1RSB picture, configurations at energy level $\ell$ are arranged in separate clusters. Each cluster has a vanishing entropy at zero temperature and the number of clusters is exponentially
large in $N$, with a rate given by the so-called \emph{complexity function} $\Sigma(\ell)$ (which implicitly depends on $\lambda$), and with \emph{intensive} energy $\ell= \mathcal{L} / N$. These quantities are linked via the Legendre relation $s\Phi(s) = \sup_{\ell} \Sigma(\ell) - \ell s$.

For large $s$, we obtain the lowest energy configurations, which are indeed the solutions of the original optimization problem \eqref{eq: minimization formulation}. However, one cannot directly take the limit 
 $s\to\infty$. In fact, for sufficiently large values of $s$ the complexity becomes negative (i.e. $e^{N\Sigma(\ell)}\ll 1$), meaning that the 1RSB picture is describing clusters whose presence in a sample is a rare event. The physical prescription, which is part of the extremization condition for 
\eqref{eq:free_energy_1rsb}, is to take $s^\star={\rm sup}\, {s}$ s.t. $ \Sigma(\ell^\star)\geq0$.
The complexity can be computed via the Legendre identity  $\Sigma(\ell(s)) = -s^2\partial_s \Phi(s)$.

As for the RS free energy and AMP, we can link the 1RSB computation with our ASP algorithms, by noticing that the free energy saddle-point equations correspond to the ASP SE from Eq.~\ref{eq: ASP SE}. Indeed, the set of saddle-point equations reads
\begin{align}
    \hat{q}_0&=\frac{\alpha}{2}\frac{\E[{x_o}^2]_{{x_o}}-2m+q_0}{(1+V_1+{s}V_0)^2},\\
    \hat{m}&=A_1-sA_0,\nonumber\\
    V_0&=q_0 -\!2{\rm I\!E}\Big[\partial^2_{{A_1}} \phi^{\rm 1RSB}_{in} \Big(\sqrt{\hat{q}_0}{z}\!+\!(A_1\!-\!sA_0){{x}_o},{A}_0,{A}_1,s,\lambda\Big)\Big]_{{{x}_o},{z}} , \nonumber\\
    V_1+sV_0&={\rm I\!E}\Big[\partial^2_{B} \phi^{\rm 1RSB}_{in} \Big(\sqrt{\hat{q}_0}{z}=\!+\!(A_1\!-\!sA_0){{x}_o},{A}_0,{A}_1,s,\lambda\Big)\Big]_{{{x}_o},{z}} , \nonumber\\
     {A}_0&=\frac{\alpha}{s}\left(\frac{1}{1+V_0}-\frac{1}{1+{V}_1+s{V}_0}\right) \nonumber\, ,\\
    {A}_1-s{A}_0&=\frac{\alpha}{1+{V}_1+s{V}_0}
    \nonumber, \\
    m&={\rm I\!E}\Big[x_o\partial_{B} \phi^{\rm 1RSB}_{in} \Big(\sqrt{\hat{q}_0}{z}\!+\!(A_1\!-\!sA_0){{x}_o},{A}_0,{A}_1,s,\lambda\Big)\Big]_{{{x}_o},{z}} , \nonumber\\
    q&={\rm I\!E}\Bigg\{\Big[\partial_{B} \phi^{\rm 1RSB}_{in} \Big(\sqrt{\hat{q}_0}{z}\!+\!(A_1\!-\!sA_0){{x}_o},{A}_0,{A}_1,s,\lambda\Big)\Big]^2\Bigg\}_{{{x}_o},{z}} . \nonumber
\end{align}
and by simply reinjecting the equations for $\hat{q}_0$, $\hat{m}$, $V_0$, and $V_1$ into the remaining ones we recover the ASP state evolution without time indices.

For sufficiently low values of $\lambda$, we recall that the saddle point equations for the 1RSB free energy admit a stable fixed point that can be reached by iteration starting from $m=0$ (uninformed initialization).  This fixed point is present for any signal density $\rhoo$ and any compression rate $\alpha$. As $\lambda$ goes to zero, two regimes can be observed, corresponding respectively to a low compression ($\alpha>\alpha_{c,{\rm 1RSB }}(\rhoo)$) regime where the fixed point describes the original signal,
and a high compression ($\alpha<\alpha_{c,{\rm 1RSB }}(\rhoo)$) where the fixed point describes configurations at large density. Within the replica formalism, one can 
compute the expected reconstruction error $\epsilon_\text{rec}(\lambda)=\lim_{N\to\infty}\E\left[\langle\Vert\bxo- \bx\Vert_2^2\rangle\right]$. 
For $\alpha<\alpha_{c,{\rm 1RSB }}(\rhoo)$, we observe that $\epsilon_\text{rec}(\lambda)\to 0$ as $\lambda\to0$.
For $\alpha>\alpha_{c,{\rm 1RSB }}(\rhoo)$ instead,
$\epsilon_\text{rec}(0^+) \neq 0$. In this regime, 
where we don't have perfect recovery, we still observe 
a partial alignment of the reconstructed signal with
the true one (weak recovery, $m>0$). The configurations described by this 1RSB solution are also well separated from each other ($V_0 > 0$ in the replica formalism),
satisfy the measurement constraints ($e=0$), but are
not as sparse as the true signal ($\rho>\rhoo$).

\begin{figure}
      \centering     
      \includegraphics[width=0.43\textwidth]{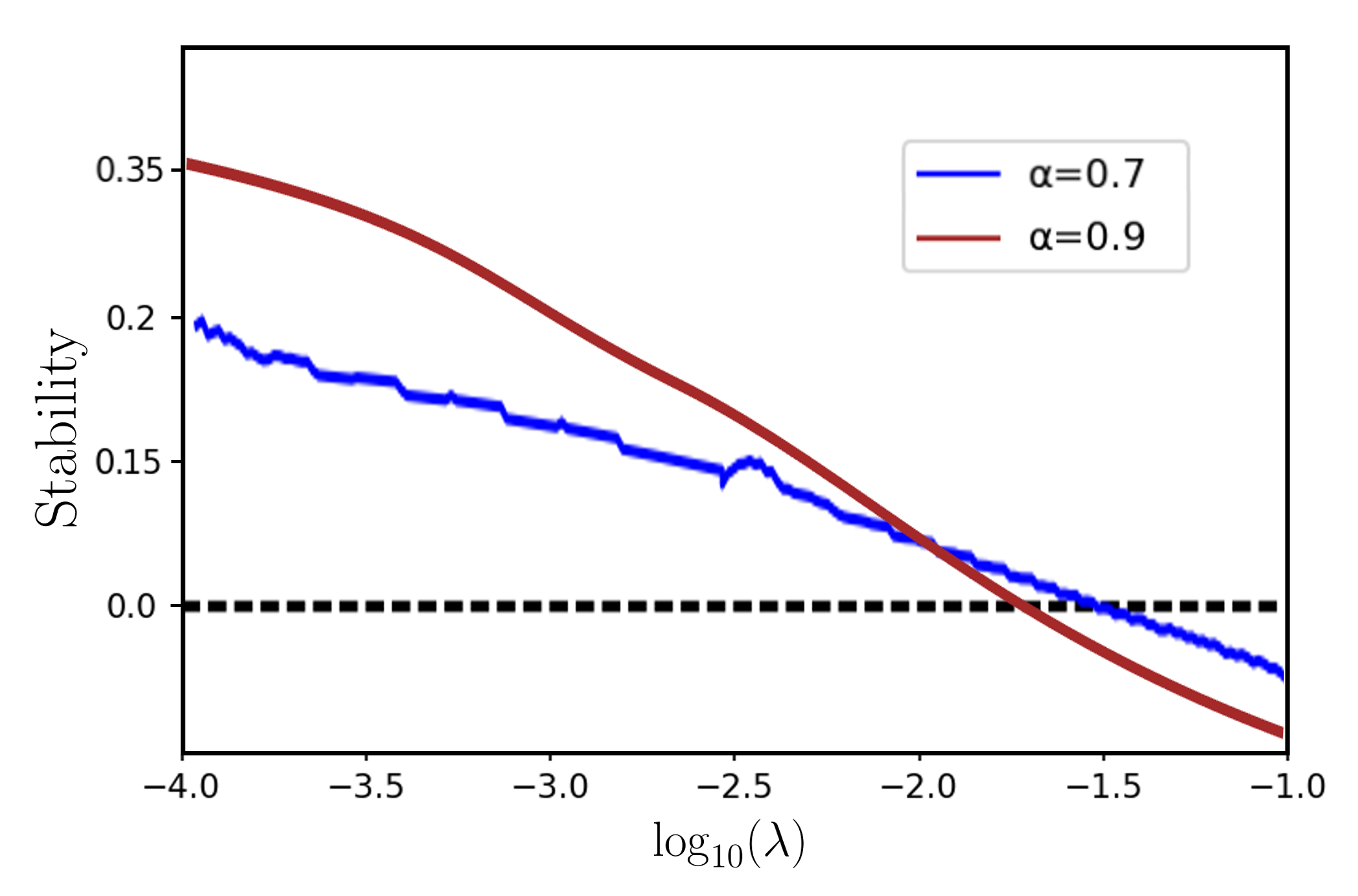}
\caption{Stability condition (type I as defined in~\cite{Antenucci2019}) for the 1RSB saddle point both in the regimes where it yields signal recovery ($\alpha=0.9$) and where it does not ($\alpha=0.7$). Regardless of the regime, we observe that the 1RSB saddle point is stable when $\lambda$ is low enough, i.e. the curves lay above zero.}
      \label{fig: 1RSB stability}
  \end{figure}

Finally, Fig.~\ref{fig: 1RSB stability} shows the stability of the 1RSB solution against further steps of replica symmetry breaking \cite{Antenucci2019}. In particular, we probed the stability of the 1RSB $Ansatz$ by perturbing the entries of the overlap matrix $Q$, allowing it to take the two-steps replica symmetry breaking (2RSB) form. In this context two types of perturbation are possible. Either we can further break the symmetry for replicas lying in the same basin - this is called $states$ $splitting$ or type I instability- or we can further break the symmetry between replicas lying in different basins - this is called $states$ $aggregation$ or type II instability. Following primarily \cite{Antenucci2019}, but also \cite{Montanari2003,Krzakala04,Crisanti2006}, we plotted in Fig.~\ref{fig: 1RSB stability} the quantity accounting for the type I instability. When this quantity is positive the system is stable with respect to the associated 2RSB perturbation. We can see that, in both regimes and for low enough regularization prefactor, the 1RSB saddle-point is stable with respect to the type I perturbation since the plotted quantity is positive. Regarding the type II perturbation, this quantity suffers from the same non-differentiable functions as the RS stability criteria analysed in App.~\ref{sec: RS stability}. We find that the diverging integral appearing in the type II perturbation is the same as in the RS stability criteria. We mentioned in App.~\ref{sec: RS stability} that this integral, when analysed properly, indicates that the system is stable towards this perturbation. Therefore, being stable towards the type I and II perturbations our problem does not necessarily call for further replica symmetry breaking in order to compute the correct free energy (in the limit $\lambda \rightarrow 0^+$).

\section{Conclusion}
This paper follows the study initiated in \cite{Kabashima2009} on the implementation of a $\lzero$-norm penalty in a compressed sensing setting. 
In the first part of the discussion, we showed how one can in practice access the low-energy configurations of the problem with a survey propagation algorithm, following the approach of \cite{Antenucci2019,Lucibello19a}. We also showed how the recovery algorithm can be simplified to an approximate message passing scheme that we called ${\rm ASP_o}$. In particular, we highlighted how the two algorithms outperform the LASSO iteration, recovering the sparse signal for a wider range of compression rates.

We then derived a statistical physics analysis of the minimization problem with cost function $\mathcal{L}(\bx,\by,\bF,\lambda)=\Vert\by-\bF\bx\Vert_2^2+\lambda\Vert \bx \Vert_0$. We showed that the minima of the loss are arranged in a clustered (1RSB) structure. We also showed how two regimes emerge as the $\lzero$ norm penalty is switched off: an ``easy" phase at low compression rate, where the 1RSB clusters collapse on the true signal (when $\lambda\rightarrow 0^+$), and a ``hard" phase at high compression rate, where the clusters remain distant from each other and from the signal. We also showed that this 1RSB saddle-point remains stable for sufficiently low $\lambda$ and therefore further steps of replica symmetry breaking are not necessary.

An open question remains on the performance of these algorithms when noise is added to the measurement protocol. Additionally, while the algorithm's dynamics can be rigorously tracked by state evolution, a rigorous analysis of the replica solution would be desirable. Finally, it would also be interesting to apply the ASP algorithms to other inference problems and to connect the hard phase to the overlap-gap property of \cite{gamarnik2021overlap}.

\section{Acknowledgement}
We thank M. Mézard, A. Bereyhi, and P. Urbani for very helpful discussions.
We also acknowledge funding from the ERC under the European Union’s Horizon 2020 Research and Innovation Program Grant Agreement 714608-SMiLe as well as from the Swiss National Science Foundation grant SNFS OperaGOST, $200021\_200390$.

\appendix

\section{Finite temperature RS and 1RSB free energies}
\label{app: finite temp free energy}
We detail in this appendix the form of the RS and 1RSB free energies before taking the zero-temperature limit. If we plug the replica symmetric $Ansatz$ \cite{Kabashima2009,Kabashima2010,Krzakala2012_1,Krzakala2012_2} we obtain the free energy
\begin{eqnarray}
\label{eq: RS free energy}
&&\frac{1}{N}{\rm I\!E}\left[\log[{Z}(\bm{y},\bm{F},\lambda)]\right]_{\bm{y},\bm{F}}=\underset{Q,\tilde{Q},q,\tilde{q},m,\tilde{m}}{\rm extr}\Bigg\{-\tilde{m}m+\frac{\tilde{Q}Q+\tilde{q}q}{2}+{\rm I\!E}\Big[ \tilde{\phi}^{\rm RS}_{in}(z\sqrt{\tilde{q}}+\tilde{m}x_o,\tilde{Q}+\tilde{q},\lambda)\Big]_{{x_o}, z} \\
&&\hspace{6.5cm}+{\rm I\!E}\Big[ \tilde{\phi}^{\rm RS}_{out}\Big(z\sqrt{{q}},Q\!-\!q,z\sqrt{{\rm I\!E}\Big[ {x_o}^2\Big]_{{x_o}}\!\!-\!\frac{m^2}{q}}+z'\frac{m}{\sqrt{q}}\Big)\Big]_{z, z'}\Bigg\}\, ,\nonumber
\end{eqnarray}
with
\begin{eqnarray}
\tilde{\phi}^{\rm RS}_{in}(B,A,\lambda)&=&\log\int dx\, e^{-\frac{A}{2}x^2+Bx-\beta\lambda\vert\vert x\vert\vert_o}\, ,\\
\tilde{\phi}^{\rm RS}_{out}(w,V,y)&=&\log{\rm I\!E}\Big[ e^{-\beta(y-\sqrt{V}z-w)^2}\Big]_{z}\, ,
\end{eqnarray}
 where ${\rm I\!E}[\cdot]_{z}$ is a Gaussian integration measure with zero mean and variance one (the same will apply for ${\rm I\!E}[\cdot]_{z'}$ later in this appendix). Moreover in the previous equation ${\rm I\!E}[\cdot]_{x_o}$ indicates the average over one entry of the signal. Finally the $\beta\rightarrow+\infty$ limit can be taken using the reparametrization 
\begin{eqnarray}
\label{eq: RS zero-temperature rescaling }
A&=&\beta^{-1}(\tilde{q}+\tilde{Q})\, ,\quad V=\beta(Q-q) \\
\hat{q}&=&\beta^{-2}\tilde{q}\, ,\hspace{1.161cm}\,\, \hat{m}=\beta^{-1} \tilde{m}\, .\nonumber
\end{eqnarray}

Then if we consider the 1-step replica symmetry breaking. Following \cite{Antenucci2019,Obuchi2018,Bereyhi2017} we obtain the free energy 
\begin{eqnarray}
&&\frac{1}{N}{\rm I\!E}\left[\log[{Z}(\bm{y},\bm{F},\lambda)]\right]_{\bm{y},\bm{F}}=\underset{Q,\tilde{Q},q_1,\tilde{q}_1,q_0,\tilde{q}_0,m,\tilde{m}}{\rm extr}\Bigg\{\!\!-\!\tilde{m}m+\frac{\tilde{Q}Q+(1-\tilde{s})\tilde{q}_1 q_1+\tilde{s}\tilde{q}_0 q_0}{2}\\
&&\hspace{7cm}+{\rm I\!E}\Big[ \tilde{\phi}^{\rm 1RSB}_{in}(z\sqrt{\tilde{q}_0}+\tilde{m}x_o,\tilde{Q}+\tilde{q}_1,\tilde{q}_1+\tilde{q}_0,\tilde{s},\lambda)\Big]_{{x_o}, z} \nonumber\\
&&\hspace{7cm}+{\rm I\!E}\Big[ \tilde{\phi}^{\rm 1RSB}_{out}\Big(z\sqrt{{q_0}},Q\!-\!q_1,q_1\!-\!q_0,z\sqrt{{\rm I\!E}\Big[ {x_o}^2\Big]_{{x_o}}\!\!-\!\frac{m^2}{q_0}}+z'\frac{m}{\sqrt{q_0}}\Big)\Big]_{z, z'}\Bigg\}\, ,\nonumber
\end{eqnarray}
with
\begin{align}
\tilde{\phi}^{\rm 1RSB}_{in}(B,A_0,A_1,\tilde{s},\lambda)&\!=\!\frac{1}{\tilde{s}}\log\!{\rm I\!E}\!\left[ e^{-\tilde{s}\tilde{\phi}^{\rm RS}_{in}(B+\sqrt{A_0}z,A_1,\lambda)}\right]_{z} \!,\quad\quad\!\!\\
\tilde{\phi}^{\rm 1RSB}_{out}(w,V_0,V_1,y)&\!=\!\frac{1}{\tilde{s}}\log\!{\rm I\!E}\!\left[ e^{-\tilde{s}\tilde{\phi}^{\rm RS}_{out}(w+\sqrt{V_0}z,V_1,y)}\right]_{z} \!\!.
\end{align}
In this case the $\beta\rightarrow+\infty$ limit can be taken using the reparametrization 
\begin{eqnarray}
A_1&=&\beta^{-1}(\tilde{q}_1+\tilde{Q})\, ,\quad V_1=\beta(Q-q_1)\, , \\
A_0&=&\beta^{-2}(\tilde{q}_0+\tilde{q}_1)\, ,\,\,\hspace{0.2 cm} V_0=(q_1-q_0)\, ,\nonumber\\
\hat{q}_0&=&\beta^{-2}\tilde{q}_0\, ,\quad\qquad\,\, \hspace{0.2 cm}\hat{m}=\beta^{-1} \tilde{m}\, ,\nonumber\\
{s}&=&\tilde{s}\beta\, .\nonumber
\end{eqnarray}

\section{Finite temperature AMP and ASP algorithms}
\label{app: finite temp algo}
As explained in Sec.~\ref{subsec: HT algorithms} we can derive an approximate message passing algorithm based on the finite-temperature RS computation of the free energy. Following exactly step by step \cite{Krzakala2012_1} we obtain the iteration scheme (with for example initialization $\bm{\hat{x}}^{t=0}=\bxo$, $\tilde{V}^{t=0}=0$, $\bm{\tilde{g}}^{t=0}=0$)
\begin{align}
\label{eq: AMP algo finite temp}
w_\mu^t&=\sum_{i=1}^N \bm{F}_{\mu i}\hat{x}_i^{t-1}-\tilde{g}_\mu^{t-1}\tilde{V}^{t-1}\, ,\\
\tilde{g}_\mu^t&=\frac{y_\mu-w_\mu^t}{\beta^{-1}+\tilde{V}^{t-1}}\, , \quad \mu\in{[\![} 1,M{]\!]}\, ,\nonumber\\
\tilde{A}^{t}&=\frac{\alpha}{\beta^{-1}+\tilde{V}^{t-1}}\, ,\nonumber\\
\tilde{B}^t_i&=\sum_{\mu=1}^M \bm{F}_{\mu i}\tilde{g}_\mu^{t}+\hat{x}_i^{t-1}\tilde{A}^t\, , \nonumber\\
\hat{x}_i^t&=\partial_{\tilde{B}} \phi^{\rm RS}_{in}(\tilde{B}_i^t,\tilde{A}^{t},\lambda)\, , \quad i\in{[\![} 1,N{]\!]}\, ,\nonumber\\
\tilde{V}^t&=\frac{1}{N}\sum_{i=1}^N \partial^2_{\tilde{B}} \phi^{\rm RS}_{in}(\tilde{B}_i^t,\tilde{A}^{t},\lambda)\; . \nonumber
\end{align}
Equivalently with the computation of the RS saddle-point we can take the $\beta\rightarrow +\infty$ limit using the rescaling:
\begin{eqnarray}
A&=&\beta^{-1}\tilde{A}\, , \quad V=\beta\tilde{V}\, ,\\
\bm{B}&=&\beta^{-1}\bm{\tilde{B}}\, ,
\quad \bm{g}=\beta^{-1}\bm{\tilde{g}}\;. \nonumber
\end{eqnarray}
Thus the algorithm from Eq.~(\ref{eq: AMP algo}) simply corresponds to Eq~(\ref{eq: AMP algo finite temp}) after applying this rescaling. Its form is in fact more compact as we  used the rewriting
\begin{eqnarray}
\bm{z}^t=\bm{y}-\bm{w}^t\, ,\quad \frac{\bm{B}^t}{A^t}&=&\frac{\bm{F}^{\intercal}\bm{z}^{t-1}}{\alpha}+\bm{\hat{x}}^{t-1}\, .
\end{eqnarray}

We can in fact derive quite similarly the approximate survey propagation algorithm to probe for potential 1RSB fixed points of our problem.
Following \cite{Krzakala2012_2,Antenucci2019} we obtain (with for example the initialization $\bm{\hat{x}}^{t=0}=0$, $\tilde{V}_1^{t=0}=\tilde{V}_0^{t=0}=0$, $\bm{\tilde{g}}^{t=0}=0$)
\begin{align}
\label{eq: ASP algo finite temp}
    w_\mu^t&=\sum_i F_{\mu i}\hat{x}_i^t-\tilde{g}_\mu^{t-1}(\tilde{s}\,\tilde{V}_0^{t-1}+\tilde{V}_1^{t-1}) \, ,\\
    \tilde{g}_\mu^t&=\partial_{w}\tilde{\phi}_{out}(w_\mu^t,\tilde{V}_0^t,\tilde{V}_1^t,y_\mu)\, , \quad \mu\in{[\![} 1,M{]\!]}\, ,\nonumber\\
    \tilde{A}_0^{t+1}&=\frac{\sum_\mu\left[(\partial^2_{w}-2\partial^2_{\tilde{V}_1})\tilde{\phi}^{\rm 1RSB}_{out}(w^t_\mu,\tilde{V}_0^t,\tilde{V}_1^t,y_\mu)-(\tilde{g}_\mu^t)^2\right]}{N(\tilde{s}-1)} \nonumber\, ,\\
    \tilde{A}_1^{t+1}&=\frac{\sum_\mu\left[(\partial^2_{w}-2\tilde{s}\partial^2_{\tilde{V}_1})\tilde{\phi}^{\rm 1RSB}_{out}(w^t_\mu,\tilde{V}_0^t,\tilde{V}_1^t,y_\mu)-\tilde{s}(\tilde{g}_\mu^t)^2\right]}{N(\tilde{s}-1)} \nonumber\, ,\\
    \tilde{B}_i^{t+1}&=\sum_\mu F_{\mu 
    i}\tilde{g}_\mu^t+\hat{x}_i^t(\tilde{A}_1^{t+1}-\tilde{s}\tilde{A}_0^{t+1}) \nonumber\, ,\\
    \hat{x}_i^t &=\partial_{\tilde{B}} \tilde{\phi}^{\rm 1RSB}_{in}(\tilde{B}_i^t,\tilde{A}_0^t,\tilde{A}_1^t,\tilde{s},\lambda)\, , \quad \mu\in{[\![} 1,N{]\!]}\, ,\nonumber\\
    \tilde{V}_0^t&=\frac{\sum_i\left[(\partial^2_{\tilde{B}}+2\partial^2_{\tilde{A}_1})\tilde{\phi}^{\rm 1RSB}_{in}({\tilde{B}}_i^t,\tilde{A}_0^t,\tilde{A}_1^t,\tilde{s},\lambda)+(\hat{x}_i^t)^2\right]}{N(\tilde{s}-1)} \nonumber\, ,\\
    \tilde{V}_1^t&=\partial^2_{\tilde{B}} \tilde{\phi}^{\rm 1RSB}_{in}(\tilde{B}_i^t,\tilde{A}_0^t,\tilde{A}_1^t,\tilde{s},\lambda)-\tilde{s}\,\tilde{V}_0^t \nonumber\, .
\end{align}

As previously preformed with the 1RSB free energy we can take the limit $\beta\rightarrow +\infty$ with the rescaling
\begin{eqnarray}
A_1&=&\beta^{-1}\tilde{A}_1\, , V_1=\beta\tilde{V}_1\, , \\
A_0&=&\beta^{-2}\tilde{A}_2\, ,\hspace{0.2 cm}s=\tilde{s}\beta\, ,\nonumber\\
\bm{B}&=&\beta^{-1}\bm{\tilde{B}}\, ,\, \hspace{0.2 cm}\bm{g}=\beta^{-1} \bm{\tilde{g}}\,  .\nonumber
\end{eqnarray}
The ASP algorithm derived with this rescaling can be found in Eq.~(\ref{eq: ASP algo}).

\section{Stability of the RS solution}
\label{sec: RS stability}
In this section we will discuss the stability of the replica symmetric fixed point. To do so we look at the 1RSB free energy and its landscape around the point:
\begin{align}
    &q=q_{\rm RS},\, m=m_{\rm RS}, \, V_1=V_{\rm RS},\,V_0=0\, ,\\
    &\hat{q}=\hat{q}_{\rm RS},\, \hat{m}=\hat{m}_{\rm RS}, \, A_1=A_{\rm RS},\,A_0=0 \, .\nonumber
\end{align}
More particularly we check if the RS saddle-point is a stable fixed point of the 1RSB free energy. Unlike most usual RS stability check, our case involves a 1RSB free energy that is non-differentiable around $A_0=0$. Indeed, this feature becomes apparent after developing the function ${\phi}^{\rm 1RSB}_{in}(B,A_1,A_0,\lambda)$ as
\begin{align}
  &  {\phi}^{\rm 1RSB}_{in}(B,A_1,A_0,\lambda)=\frac{1}{s}\log\left\{\frac{e^{-s\lambda+\frac{sB^2}{A_1-sA_0}}\sqrt{A_1}}{2\sqrt{A_1-sA_0}}\right.\\
  &\hspace{0.9cm}\times\left[{\rm erfc}\left(\frac{\sqrt{2\lambda}(A_1-sA_0)+\sqrt{A_1}B}{\sqrt{2A_0(A_1-sA_0)}}\right)\right.\nonumber\\
  &\left.\hspace{1.2cm}+{\rm erfc}\left(\frac{\sqrt{2\lambda}(A_1-sA_0)+\sqrt{A_1}B}{\sqrt{2A_0(A_1-sA_0)}}\right)\right]\nonumber\\
  &\hspace{0.6cm}
  \left.+\frac{1}{2}{\rm erfc}\left(\frac{B-\sqrt{2\lambda A_1}}{\sqrt{2A_0}}\right)-\frac{1}{2}{\rm erfc}\left(\frac{B+\sqrt{2\lambda A_1}}{\sqrt{2A_0}}\right)\right\}\nonumber\, ,
\end{align}
${\rm erfc}(.)$ being the complementary error function.  In this case the function is non-differentiable around the RS saddle-point as it contains $\sqrt{A_0}$  terms. Thus, we numerically checked that the perturbation along $A_0$ actually unveils an infinitely stable direction, i.e.
\begin{align}
 \E\Big[\partial^2_{A_0} {\phi}^{\rm 1RSB}_{in}(z\sqrt{\hat{q_0}}+\hat{m}x_o,A_1,A_0,\lambda)\Big]_{{x_o}, z} \Big\vert_{{A_0\rightarrow 0^+}}  =-\infty
\end{align}
Computing this quantity via the Anderson-Thouless stability criteria ~\cite{Antenucci2019} would should that this direction is unstable, while this more careful check shows that is not the case. This finally indicates that the RS saddle point is stable towards further replica symmetry breaking.

\section{Stability of the 1RSB solution (type I)}
\label{app: stability type I}
To obtain the type I stability criterion (for the 1RSB computation) one possibility is to study the behavior of the associated survey propagation algorithm under perturbation. As a reminder the approximate survey propagation algorithm is 
\begin{align}
    w_\mu^t&=\sum_i F_{\mu i}\hat{x}_i^t-{g}_\mu^{t-1}({s}\, {V}_0^{t-1}+ {V}_1^{t-1}) \, ,\\
     {g}_\mu^t&=\frac{y_\mu-w_\mu^t}{1+{V}_1^{t-1}+s{V}_0^{t-1}}\, , \quad \mu\in{[\![} 1,M{]\!]}\, ,\nonumber\\
    {A}_0^{t+1}&=\frac{\alpha}{s}\left(\frac{1}{1+V_0^{t-1}}-\frac{1}{1+{V}_1^{t-1}+s{V}_0^{t-1}}\right) \nonumber\, ,\\
    {A}_1^{t+1}&=\frac{\alpha}{1+{V}_1^{t-1}+s{V}_0^{t-1}}+s{A}_0^{t+1}
    \nonumber\, ,\\
    {B}_i^{t+1}&=\sum_\mu F_{\mu 
    i}{g}_\mu^t+\hat{x}_i^t({A}_1^{t+1}-{s}{A}_0^{t+1}) \nonumber\, ,\\
    \hat{x}_i^t &=\partial_{{B}} \phi^{\rm 1RSB}_{in}({B}_i^t,{A}_0^t,{A}_1^t,s,\lambda)\, , \quad \mu\in{[\![} 1,N{]\!]}\, ,\nonumber\\
    {V}_0^t&=\frac{\sum_i\left[-2\partial^2_{{A}_1}\phi^{\rm 1RSB}_{in}({B}_i^t,{A}_0^t,{A}_1^t,s,\lambda)+(\hat{x}_i^t)^2\right]}{N} \nonumber\, ,\\
    {V}_1^t&=\partial^2_{{B}} \phi^{\rm 1RSB}_{in}({B}_i^t,{A}_0^t,{A}_1^t,s,\lambda)-{s}\,{V}_0^t \nonumber\, .
\end{align}

To get the relevant stability in our problem we perturb $\Vec{B}^t$ as $\Vec{B}^t\rightarrow \Vec{B}_i^{t,(0)}+\vec\delta {\epsilon}$, where $\vec\delta {\epsilon}$ is an i.i.d. infinitesimal vector sampled from $\mathcal{N}(0,\epsilon)$. We then compute the next iteration at leading order in $\epsilon$. More precisely we have
\begin{eqnarray}
&&\hat{x}_i^t \approx\hat{x}_i^{t,(0)}+\partial^2_B \phi_{in}(B_i^t,A_0^t,A_1^t,\lambda)\delta {\epsilon}_i \\
&\implies& w_\mu^t\approx w_\mu^{t,(0)}+\sum_i F_{\mu i}\partial^2_B \phi_{in}(B_i^t,A_0^t,A_1^t,\lambda)\delta {\epsilon}_i\nonumber\\
&\implies& g_\mu^t \approx g_\mu^{t,(0)}-\frac{\sum_i F_{\mu i}\partial^2_B \phi_{in}(B_i^t,A_0^t,A_1^t,\lambda)\delta {\epsilon}_i}{1+V_1^t+sV_0^t} \nonumber\\
&\implies& B_i^{t+1} \approx B_i^{t+1,(0)}-\frac{\sum_{\mu,j} F_{\mu i}F_{\mu j}\partial^2_B \phi_{in}(B_j^t,A_0^t,A_1^t,\lambda)\delta {\epsilon}_j}{1+V_1^t+sV_0^t}+\partial^2_B \phi_{in}(B_i^t,A_0^t,A_1^t,\lambda)(A_1^{t+1}-sA_0^{t+1})\delta {\epsilon}_i \nonumber
\end{eqnarray}
and therefore the new perturbation for $B_i^{t+1}$ is
\begin{eqnarray}
\delta {\epsilon}^{new}_i&=&-\frac{\sum_{\mu,j} F_{\mu i}F_{\mu j}\partial^2_B \phi_{in}(B_j^t,A_0^t,A_1^t,\lambda)\delta {\epsilon}_j}{1+V_1^t+sV_0^t}\\
&&+\partial^2_B \phi_{in}(B_i^t,A_0^t,A_1^t,\lambda)(A_1^{t+1}-sA_0^{t+1})\delta {\epsilon}_i \, .\nonumber
\end{eqnarray}
Its averaged $l_2$ norm is
\begin{eqnarray}
&&\Big\langle\sum_i (\delta {\epsilon}^{new}_i)^2\Big\rangle_{\vec\delta {\epsilon}}\!=\!\sum_i\Bigg[\sum_{\mu,\mu'}\sum_{j,j'}\frac{F_{\mu i}F_{\mu j} F_{\mu' i}F_{\mu' j'}}{(1+V^t_1+yV^t_0)^2}\partial^2_B \phi_{in}(B_j^t,A_0^t,A_1^t,\lambda)\partial^2_B \phi_{in}(B_{j'}^t,A_0^t,A_1^t,\lambda) \langle\delta {\epsilon}_j\delta {\epsilon}_{j'}\rangle_{\vec\delta {\epsilon}}\\
&&\hspace{3.8cm}-2\alpha\sum_{\mu,j} \frac{F_{\mu i}F_{\mu j}}{{(1+V^t_1+yV^t_0)^2}}\partial^2_B \phi_{in}(B_j^t,A_0^t,A_1^t,\lambda)\partial^2_B \phi_{in}(B_i^t,A_0^t,A_1^t)\langle\delta {\epsilon}_j\delta{\epsilon}_i\rangle_{\vec\delta {\epsilon}}\nonumber\\
&&\hspace{3.8cm}+\frac{\langle(\alpha\partial^2_B \phi_{in}(B_i^t,A_0^t,A_1^t,\lambda)\delta{\epsilon}_i)^2\rangle_{\vec\delta {\epsilon}}}{{(1+V^t_1+yV^t_0)^2}}\Bigg]\, ,\nonumber
\end{eqnarray}
where $\langle . \rangle_{\vec\delta {\epsilon}}$ indicates the average over all entries $\delta\epsilon_i$
The sums in the previous equation are dominated by two terms, one corresponding to $j=j'=i$ and another one corresponding to $\mu=\mu'; j=j'$. Considering this  we finally obtain
\begin{eqnarray}
\frac{1}{N}\left\langle\sum_i (\delta {\epsilon}^{new}_i)^2\right\rangle_{\vec\delta {\epsilon}}&\!=\!&\frac{\alpha\epsilon^2 \sum_i(\partial^2_B \phi_{in}(B_i^t,A_0^t,A_1^t))^2}{N(1+V^t_1+yV^t_0)^2}\, .\qquad\quad
\end{eqnarray}
The algorithm is considered unstable when the perturbation grows with respect to the $l_2$-norm, in other words when we have
\begin{eqnarray}
&&\frac{1}{N}\left\langle\sum_i (\delta {\epsilon}^{new}_i)^2\right\rangle_{\vec\delta {\epsilon}}>\epsilon^2\, \\
&\implies& 1-\frac{\alpha \sum_i(\partial^2_B \phi_{in}(B_i^t,A_0^t,A_1^t))^2}{N(1+V^t_1+yV^t_0)^2}<0\, . \nonumber
\end{eqnarray}
\nocite{*}

\bibliography{compressed_sensing}

\end{document}